\def\draft{0}
\def\llncs{0}
\def\ieee{1}
\def\anon{0}

\ifnum\llncs=1
\documentclass[runningheads]{llncs}
\usepackage{amssymb,amsmath,cite,url}
\else
\ifnum\ieee=1
\documentclass[conference,letterpaper]{IEEEtran}
\usepackage{amssymb,amsmath,cite,url}
\else
\documentclass[11pt,letterpaper]{article}
\usepackage{amsfonts, amsmath, cite, url,amsthm,amssymb}
\usepackage{fullpage}
\fi

\usepackage{times}
\usepackage{booktabs}
\usepackage{siunitx}

\usepackage{algorithmicx}
\usepackage{multirow}
\usepackage[table]{xcolor}
\usepackage{colortbl}
\usepackage{graphicx}
\usepackage[para]{threeparttable}

\usepackage{makecell}
\usepackage{boldline}
\setcellgapes{3pt}

\title{{{\sc gazelle}: A Low Latency Framework for Secure Neural Network Inference}}

\ifnum\anon=0
\author{\IEEEauthorblockN{Chiraag Juvekar}
\IEEEauthorblockA{MIT MTL\\ \texttt{chiraag@mit.edu}}
\and
\IEEEauthorblockN{Vinod Vaikuntanathan}
\IEEEauthorblockA{MIT CSAIL\\ \texttt{vinodv@csail.mit.edu}}
\and
\IEEEauthorblockN{Anantha Chandrakasan}
\IEEEauthorblockA{MIT MTL\\ \texttt{anantha@mtl.mit.edu}}
}
\else
\author{\IEEEauthorblockN{\Large \sc [Anonymous Submission]}
}
\fi

\date{\today}

\newcommand{\remove}[1]{}
\newcommand{\ignore}[1]{}

\usepackage{color}
\definecolor{DarkBlue}{RGB}{0,0,150}
\definecolor{DarkGreen}{RGB}{0,150,0}
\usepackage[colorlinks,linkcolor=DarkBlue,citecolor=DarkBlue]{hyperref}

\ifnum\llncs=0

\fi

\renewcommand{\baselinestretch}{0.997}

\def\vecb{\vc{b}}

\def\vecv{\vc{v}}
\def\vecw{\vc{w}}
\def\vecx{\vc{x}}
\def\vecy{\vc{y}}

\def\q2{\lfloor q/2 \rceil}

\newcommand{\mx}[1]{\mathbf{{#1}}}
\newcommand{\vc}[1]{\mathbf{{#1}}}

\newcommand{\out}{{\rm{out}}}

\newcommand{\pahename}{\mathsf{PAHE}}

\newcommand{\paheenc}{\mathsf{PAHE.Enc}}
\newcommand{\pahedec}{\mathsf{PAHE.Dec}}
\newcommand{\paheeval}{\mathsf{PAHE.Eval}}

\newcommand{\vecr}{\mathbf{r}}
\newcommand{\vecu}{\mathbf{u}}

\newcommand{\KeyGen}{\mathsf{KeyGen}}

\ifnum \draft=0
\newcommand{\authnote}[3]{}
\else
\newcommand{\authnote}[3]{\textcolor{#3}{[{\footnotesize {\bf #1:} { {#2}}}]}}
\fi

\newcommand{\vnote}[1]{\authnote{V}{#1}{DarkGreen}}

\newcommand{\absnewline}{\ifnum\full=0 \\ \fi}

\def\mgp[#1]{\mx{G}_{#1}}

\def\mgip[#1]{\mx{G}_{#1}^{-1}}

\def\mcit[#1]{\mci[#1]^T}
\def\mci[#1]{\mx{C}_{#1}}

\def\KeyGen{\mathsf{KeyGen}}

\def\matG{\mathbf{G}}

\def\KeyGen{\OKeyGen}

\newcounter{hybridcount}
\newcounter{prevhybridcount}
\newcounter{nexthybridcount}

\def\matW{\mathbf{W}}

\def\beginM{\left[\begin{matrix}}
	\def\endM{\end{matrix}\right]}

\def\m0{\mx{0}}

\def\mgp{\matG_p}

\def\gazelle{{\sc gazelle}}
\def\NTT{\mathsf{NTT}}

\def\overmlt{\eta_{\mathsf{mult}}}
\def\etarot{\eta_{\mathsf{rot}}}

\def\fc{\mathsf{FC}}
\def\conv{\mathsf{Conv}}

\def\KeyGen{\mathsf{KeyGen}}
\def\Encrypt{\mathsf{Encrypt}}
\def\Decrypt{\mathsf{Decrypt}}

\def\SIMDadd{\mathsf{SIMDAdd}}
\def\SIMDmult{\mathsf{SIMDScMult}}
\def\PermKeyGen{\mathsf{PermKeyGen}}
\def\Perm{\mathsf{Perm}}
\def\PermDecomp{\mathsf{PermDecomp}}
\def\PermAuto{\mathsf{PermAuto}}

\def\CostAdd{\mathsf{CostAdd}}
\def\CostMult{\mathsf{CostMult}}

\def\wrelin{\mathsf{w_{relin}}}
\def\wpt{\mathsf{w_{pt}}}

\makeatletter
\newcommand{\thickhline}{%
    \noalign {\ifnum 0=`}\fi \hrule height 1pt
    \futurelet \reserved@a \@xhline
}
\newcolumntype{"}{@{\hskip\tabcolsep\vrule width 1pt\hskip\tabcolsep}}
\makeatother

\begin{document}
\maketitle
\pagenumbering{arabic}

\begin{abstract}
The growing popularity of cloud-based machine learning raises a natural question about the privacy guarantees that can be provided in such a setting. Our work tackles this  problem in the context where a client wishes to classify private images using a convolutional neural network (CNN) trained by a server. Our goal is to build efficient protocols whereby the client can acquire the classification result without revealing their input to the server, while guaranteeing the privacy of the server's neural network. 

To this end, we design \gazelle, a scalable and low-latency system for secure neural network inference, using an intricate combination of homomorphic encryption and traditional two-party computation techniques (such as garbled circuits). \gazelle\ makes three contributions. First, we design the \gazelle\ homomorphic encryption library which provides fast algorithms for basic homomorphic operations such as SIMD (single instruction multiple data) addition, SIMD multiplication and ciphertext permutation. Second, we implement the \gazelle\ homomorphic linear algebra kernels which map neural network layers to optimized homomorphic matrix-vector multiplication and convolution routines. Third, we design optimized encryption switching protocols which seamlessly convert between homomorphic and garbled circuit encodings to enable implementation of complete neural network inference. 

We evaluate our protocols on benchmark neural networks trained on the MNIST and CIFAR-10 datasets and show that \gazelle\ outperforms the best existing systems such as MiniONN (ACM CCS 2017) by $20\times$ and Chameleon (Crypto Eprint 2017/1164) by $30\times$ in online runtime. Similarly when compared with fully homomorphic approaches like CryptoNets (ICML 2016) we demonstrate {\em three orders of magnitude} faster online run-time.
\end{abstract}



\section{Introduction}
\label{sec:intro}

Fueled by the massive influx of data, sophisticated algorithms and extensive computational resources, modern machine learning has found surprising applications in such diverse domains as medical diagnosis~\cite{diabetic,skin}, facial recognition~\cite{facenet} and credit risk assessment~\cite{creditrisk}. We consider the setting of supervised machine learning which proceeds in two phases: a \textit{training} phase where a labeled dataset is turned into a model, and an \textit{inference} or classification phase where the model is used to predict the label of a new unlabelled data point. Our work tackles a class of complex and powerful machine learning models, namely \textit{convolutional neural networks} (CNN) which have demonstrated better-than-human accuracy across a variety of image classification tasks~\cite{KSH12}.


One important use-case for such CNNs is for medical diagnosis. A large hospital with a wealth of data on, say, retinal images of patients can use techniques from recent works, e.g., \cite{diabetic}, to train a convolutional neural network that takes a retinal image as input and predicts the occurrence of a medical condition called diabetic retinopathy. The hospital may now wish to make the model available for use by the whole world and additionally, to monetize the model.

The first solution that comes to mind is for the hospital to make the model available for public consumption. This is undesirable for at least two reasons: first, once the model is given away, there is clearly no opportunity for the hospital to monetize it; and secondly, the model, which has been trained on private patient data, and may reveal information about particular patients, violating their privacy and perhaps even regulations such as HIPAA.

A second solution that comes to mind is for the hospital to adopt the ``machine learning as a service'' paradigm and build a web service that hosts the model and provides predictions for a small fee. However, this is also undesirable for at least two reasons: first, the users of such a service will rightfully be concerned about the privacy of the inputs they are providing to the web service; and secondly, the hospital may not even want to know the user inputs for reasons of legal liability in case of a data breach. 

The goal of our work is to resolve this conundrum of \textit{secure neural network inference}. More concretely we aim to provide a way for the hospital and the user to interact in such a way that the user eventually obtains the prediction (without learning the model) and the hospital obtains no information about the user's input. 

Modern cryptography provides us with many tools, in particular fully homomorphic encryption and garbled circuits, that can help us address this issue. The first key take-away from our work is that both techniques have their limitations; understanding their precise trade-offs and using a combination of them judiciously in an application-specific manner helps us overcome the individual limitations and achieve substantial gains in performance. 
Thus let us begin by discussing these two techniques and their relative merits and shortcomings.


\subsubsection*{Homomorphic Encryption} Fully Homomorphic Encryption, or FHE, is an encryption method that allows anyone to compute an arbitrary function $f$ on an encryption of $x$, without decrypting it and without knowledge of the private key~\cite{RAD78,Gentry09,BV11lwe}. As a result, one obtains an encryption of $f(x)$. Weaker versions of FHE, collectively called partially homomorphic encryption or PHE, permit the computation of a subset of all functions, typically functions that perform only additions or functions that can be computed by depth-bounded arithmetic circuits. An example of an additively homomorphic encryption (AHE) scheme is the Paillier scheme~\cite{Paillier99}. Examples of depth-bounded homomorphic encryption scheme (called leveled homomorphic encryption or LHE) are the family of {\em lattice-based} schemes such as the Brakerski-Gentry-Vaikuntanathan~\cite{BGV12} scheme and its derivatives~\cite{B12,FV12}. Recent efforts, both in theory and in practice have given us large gains in the performance of several types of PHE schemes and even FHE schemes~\cite{BGV12,GHS12,TFHEpaper,HELib,palisade,TFHElib}.

The major bottleneck for these techniques, notwithstanding these recent developments, is their \textit{computational complexity}. The computational cost of LHE, for example, grows dramatically with the number of levels of multiplication that the scheme needs to support. Indeed, the recent {\em CryptoNets} system gives us a protocol for secure neural network inference using LHE~\cite{cryptonets}. Largely due to its use of LHE, CryptoNets has two shortcomings. First, they need to change the structure of neural networks and retrain them with special LHE-friendly non-linear activation functions such as the square function (as opposed to commonly used functions such as ReLU and Sigmoid) to suit the computational needs of LHE. This has a potentially negative effect on the accuracy of these models. Secondly, and perhaps more importantly, even with these changes, the computational cost is prohibitively large. For example, on a neural network trained on the MNIST dataset, the end-to-end latency of CryptoNets is {\em $297.5$ seconds}, in stark contrast to the {\em $30$ milliseconds} end-to-end latency of \gazelle. In spite of the use of interaction, our online bandwidth per inference for this network is a mere $0.05$MB as opposed to the $372$MB required by CryptoNets. 


In contrast to the LHE scheme in CryptoNets, \gazelle\ employs, a much simpler \textit{packed additively homomorphic encryption} ($\pahename$) scheme, which we show can support very fast matrix-vector multiplications and convolutions. Lattice-based AHE schemes come with powerful features such as SIMD evaluation and automorphisms (described in detail in Section~\ref{sec:prelimsHE}) which make them the ideal tools for common linear-algebraic computations. The second key take-away from our work is that even in applications where only additive homomorphisms are required, lattice-based AHE schemes far outperform other AHE schemes such as the Paillier scheme both in computational and communication complexity.

\subsubsection*{Two Party Computation}
Yao's garbled circuits~\cite{yao86} and the Goldreich-Micali-Wigderson (GMW) protocol~\cite{GMW87} are two leading methods for the task of two-party secure computation (2PC). After three decades of theoretical and applied work improving and optimizing these protocols, we now have very efficient implementations, e.g., see~\cite{aby,spdz,libscapi,libote}. The modern versions of these techniques have the advantage of being computationally inexpensive,  partly because they rely on symmetric-key cryptographic primitives such as AES and SHA and use them in a clever way~\cite{BHKR13}, because of hardware support in the form of the Intel AES-NI instruction set, and because of techniques such as oblivious transfer extension~\cite{IKNP03,BHKR13} which limit the use of public-key cryptography to an offline reusable pre-processing phase. 

The major bottleneck for these techniques is their \textit{communication complexity}. Indeed, three recent works followed this paradigm and designed systems for secure neural network inference: the SecureML system~\cite{secureml}, the MiniONN system~\cite{minionn}, the DeepSecure system~\cite{deepsecure}. All three rely on Yao's garbled circuits.

DeepSecure uses garbled circuits alone; SecureML uses Paillier's AHE scheme to speed up some operations; and MiniONN uses a weak form of lattice-based AHE to generate so-called ``multiplication triples'' for the GMW protocol, following the SPDZ framework~\cite{spdz}. Our key claim is that understanding the precise trade-off point between AHE and garbled circuit-type techniques allows us to make optimal use of both and achieve large net computational and communication gains. In particular, in \gazelle, we use optimized AHE schemes in a completely different way from MiniONN: while they employ AHE as a pre-processing tool for the GMW protocol, we use AHE to dramatically speed up linear algebra directly.

For example, on a neural network trained on the CIFAR-10 dataset, the most efficient of these three protocols, namely MiniONN, has an online bandwidth cost of $6.2$GB whereas \gazelle has an online bandwidth cost of $0.3$GB. In fact, we observe across the board a reduction of $20$-$80\times$ in the online bandwidth per inference which gets better as the networks grow in size. In the LAN setting, this translates to an end-to-end latency of $3.6$s versus the $72$s for MiniONN. 

Even when comparing to systems such as Chameleon \cite{chameleon} that rely on trusted third-party dealers, we observe a $30\times$ reduction in online run-time and $2.5\times$ reduction in online bandwidth, while simultaneously providing a pure two-party solution, without relying on third-party dealers. (For more detailed performance comparisons with all these systems, we refer the reader to Section~\ref{sec:networks}).

\subsubsection*{(F)HE or Garbled Circuits? The Million-dollar Question} To use (F)HE and garbled circuits optimally, we need to understand the precise computational and communication trade-offs between them. Additionally, we need to (a) identify applications and the right algorithms for these applications; (b) partition these algorithms into computational sub-routines where each of these techniques outperforms the other; and (c) piece together the right solutions for each of the sub-routines in a seamless way to get a secure computation protocol for the entire application. Let us start by recapping the trade-offs between (F)HE and garbled circuits.

Roughly speaking, homomorphic encryption performs better than garbled circuits when (a) the computation has small multiplicative depth, ideally multiplicative depth $0$ meaning that we are computing a linear function; and (b) the Boolean circuit that performs the computation has large size, say quadratic in the input size. Matrix-vector multiplication (namely, the operation of multiplying a plaintext matrix with an encrypted vector) provides us with exactly such a scenario. Furthermore, the most time-consuming computations in a convolutional neural network are indeed the convolutional layers (which are nothing but a special type of matrix-vector multiplication). The non-linear computations in a CNN such as the ReLU or maxpool functions can be written as simple {\em linear-size} circuits which are best computed using garbled circuits. This analysis is the guiding philosophy that enables the design of \gazelle (For detailed descriptions of convolutional neural networks, we refer the reader to Section~\ref{sec:prelimsCNN}).

\subsubsection*{Our System}
The main contribution of this work is \gazelle, a framework for secure evaluation of convolutional neural networks. It consists of three components:
\begin{itemize}
    \item The first component is the \textit{Gazelle Homomorphic Layer} which consists of very fast implementations of three basic homomorphic operations: SIMD addition, SIMD scalar multiplication, and automorphisms (For a detailed description of these operations, see Section~\ref{sec:prelimsHE}). Our innovations in this part consist of techniques for division-free arithmetic and techniques for lazy modular reductions. In fact, our implementation of the first two of these homomorphic operations incurs only $10$-$20$x slower than the corresponding operations on plaintext, {\em when counting the number of clock cycles}.
    
    \item The second component is the \textit{Gazelle Linear Algebra kernels} which consists of very fast algorithms for homomorphic matrix-vector multiplications and homomorphic convolutions, accompanied by matching implementations. In terms of the basic homomorphic operations, SIMD additions and multiplications turn out to be relatively cheap whereas automorphisms are very expensive. At a very high level, our innovations in this part consists of several new algorithms for homomorphic matrix-vector multiplication and convolutions that minimize the expensive automorphism operations. 
    
    \item The third and final component is \textit{Gazelle Network Inference} which uses a judicious combination of garbled circuits together with our linear algebra kernels to construct a protocol for secure neural network inference. Our innovations in this part are two-fold. First, the network mapping component extracts and pre-processes the necessary garbled circuits that are required for network inference. Second, the network evaluation layer consists of efficient protocols that switch between secret-sharing and homomorphic representations of the intermediate results. 
\end{itemize} 

Our protocol also hides strictly more information about the neural network than other recent works such as the MiniONN protocol. We refer the reader to Section~\ref{sec:prelimsCNN} for more details.

\section{Secure Neural Network Inference}
\label{sec:prelimsCNN}

\begin{figure*}
    \centering
    \includegraphics[scale=0.9]{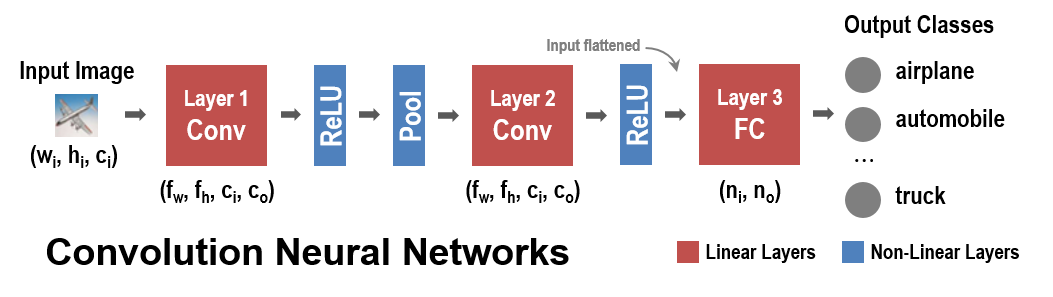}
    \caption{A CNN with two $\conv$ layers and one $\fc$ layer. ReLU is used as the activation function and a MaxPooling layer is added after the first $\conv$ layer.}
    \label{fig:cnn}
\end{figure*}

The goal of this section is to describe a clean abstraction of \textit{convolutional neural networks} (CNN)  and set up the secure neural inference problem that we will tackle in the rest of the paper. A CNN takes an input and processes it through a sequence of {\em linear} and {\em non-linear} layers in order to classify it into one of the potential classes. An example CNN is shown is Figure~\ref{fig:cnn}. 

\subsection{Linear Layers} 

The linear layers, shown in Figure~\ref{fig:cnn} in red, can be of two types: convolutional ($\conv$) layers or fully-connected ($\fc$) layers.

\subsubsection*{$\conv$ Layers}
\begin{figure}
    \centering
    \includegraphics[scale=0.6]{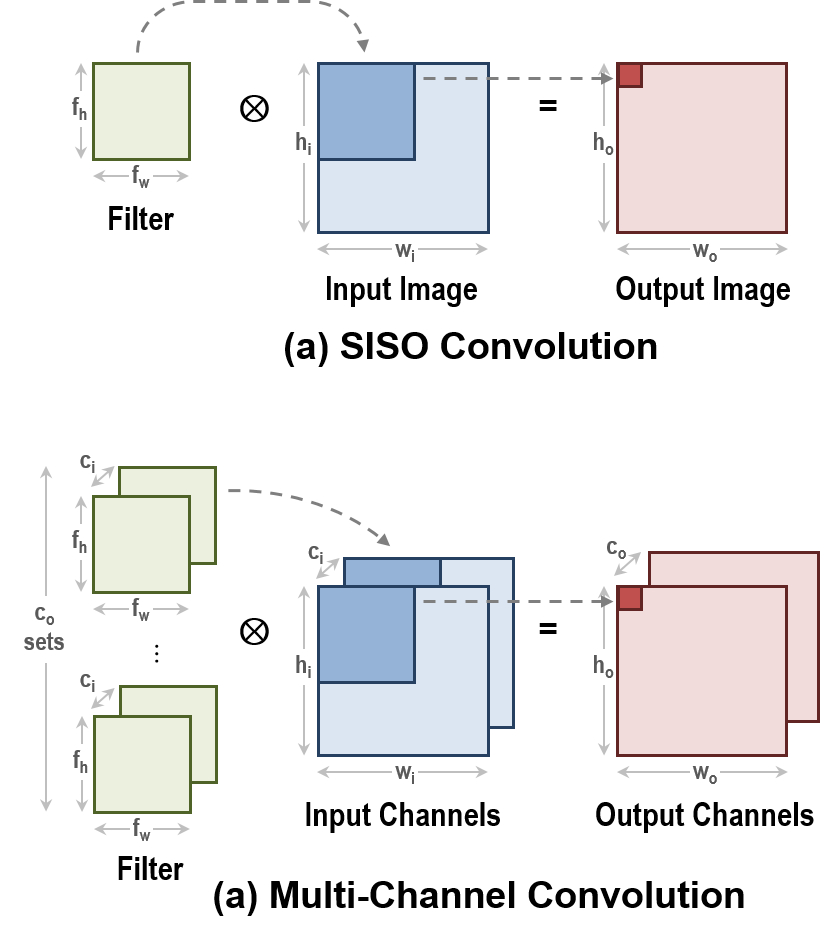}
    \caption{SISO convolutions and multi-channel $\conv$ layers}
    \label{fig:cnn-conv}
\end{figure}

We represent the input to a $\conv$ layer by the tuple ($w_i, h_i, c_i$) where $w_i$ is the image width, $h_i$ is the image height, and $c_i$ is the number of input channels. In other words, the input consists of $c_i$ many $w_i\times h_i$ images. The convolutional layer is then parameterized by $c_o$ filter banks each consisting of $c_i$ many $f_w \times f_h$ filters. This is represented in short by the tuple $(f_w, f_h, c_i, c_o)$. The computation in a $\conv$ layer can be better understood in term of simpler single-input single-output (SISO) convolutions. Every pixel in the output of a SISO convolution is computed by stepping a single $f_w\times f_h$ filter across the input image as shown in Figure~\ref{fig:cnn-conv}. The output of the full $\conv$ layer can then be parameterized by the tuple ($w_o, h_o, c_o$) which represents $c_o$ many $w_o\times h_o$ output images. Each of these images is associated to a unique filter bank and is computed by the following two-step process shown in Figure~\ref{fig:cnn-conv}: (i) For each of the $c_i$ filters in the associated filter bank, compute a SISO convolution with the corresponding channel in the input image, resulting in $c_i$ many intermediate images; and (ii) summing up all these $c_i$ intermediate images.

There are two commonly used padding schemes when performing convolutions. In the ``\textit{valid}'' scheme, no input padding is used, resulting in an output image that is smaller than the initial input. In particular we have $w_o = w_i - f_w + 1$ and $h_o = h_i - f_h + 1$. In the ``\textit{same}'' scheme, the input is zero padded such that output image size is the same as the input.

In practice, the $\conv$ layers sometimes also specify an additional pair of stride parameters ($s_w, s_h$) which denotes the granularity at which the filter is stepped. After accounting for the strides, the output image size $(w_o, h_o)$, is given by $(\lfloor(w_i - f_w + 1)/s_w\rfloor, \lfloor(h_i - f_h + 1)/s_h\rfloor)$ for \textit{valid} style convolutions and $(\lfloor w_i/s_w\rfloor, \lfloor h_i/s_h\rfloor)$ for \textit{same} style convolutions.

\subsubsection*{$\fc$ Layers}
The input to a $\fc$ layer is a vector $\vecv_i$ of length $n_i$ and its output is a vector $\vecv_o$ of length $n_o$. A fully connected layer is specified by the tuple ($\matW$, $\vecb$) where $\matW$ is $(n_o \times n_i)$ weight matrix and $\vecb$ is an $n_o$ element bias vector. The output is specified by the following transformation: $\vecv_o = \matW \vecv_i + \vecb$.

The key observation that we wish to make is that the number of multiplications in the $\conv$ and $\fc$ layers are given by $(w_o\cdot h_o \cdot c_o \cdot f_w\cdot f_h \cdot c_i)$ and $n_i\cdot n_o$, respectively. This makes both the $\conv$ and $\fc$ layer computations quadratic in the input size. This fact guides us to use homomorphic encryption rather than garbled circuit-based techniques to compute the convolution and fully connected layers, and indeed, this insight is at the heart of the much of the speedup achieved by \gazelle.

\subsection{Non-Linear Layers} 

The non-linear layers, shown in Figure~\ref{fig:cnn} in blue, consist of an activation function that acts on each element of the input separately or a pooling function that reduces the output size. Typical non-linear functions can be one of several types: the most common in the convolutional setting are max-pooling functions and ReLU functions.

The key observation that we wish to make in this context is that all these functions can be implemented by circuits that have size linear in the input size and thus, evaluating them using conventional 2PC approaches does not impose any additional asymptotic communication penalty.

 For more details on CNNs, we refer the reader to~\cite{SzeCYE17}.

\subsection{Secure Inference}
\label{sec:probstat}

In our setting, there are two parties $A$ and $B$ where $A$ holds a convolutional neural network (CNN) and $B$ holds an input to the network, typically an image. We make the distinction between the {\em architecture} of the CNN which includes the number of layers, the size of each layer, and the activation functions applied in layer, versus the {\em parameters} of the CNN which includes all the numbers that describe the convolution and the fully connected layers. 

We wish to design a protocol that $A$ and $B$ engage in at the end of which $B$ obtains the classification result, namely the output of the final layer of the neural network, whereas $A$ obtains nothing.

\subsubsection*{The Threat Model}
Our threat model is the same as in the previous works, namely the SecureML, MiniONN and DeepSecure systems and, as we argue below, leaks even less information than in these works.

To be more precise, we consider semi-honest corruptions as in \cite{deepsecure,minionn,secureml}. That is, $A$ and $B$ adhere to the software that describes the protocol, but attempt to infer information about the other party's input (the network parameters or the image, respectively) from the protocol transcript. We ask for the cryptographic standard of ideal/real security~\cite{GMR89,GMW87}.
A comment is in order about the security model.

Our protocol does not completely hide the network architecture; however, we argue that it does hide the important aspects which are likely to be proprietary. First of all, the protocol hides all the weights including those involved in the convolution and the fully connected layers. Secondly, the protocol hides the filter and stride size in the convolution layers, as well as information on which layers are convolutional layers and which are fully connected. What the protocol does reveal is the number of layers and the size (the number of hidden nodes) of each layer. At a computational expense, we are able to pad each layer and the number of layers and hide their exact numbers as well. In contrast, other protocols for secure neural network inference such as the MiniONN protocol~\cite{minionn} reveal strictly more information, e.g., they reveal the filter size. As for party $B$'s security, we hide the entire image, but not its size, from party $A$. All these choices are encoded in the definition of our ideal functionality.

\vnote{Need to describe model stealing better, postponed to v2}

\subsubsection*{Paper Organization}
The rest of the paper is organized as follows. We first describe our abstraction of a packed additively homomorphic encryption ($\pahename$) that we use through the rest of the paper. We then provide an overview of the entire \gazelle\ protocol in section \ref{sec:protocol}. In the next two sections, Section~\ref{sec:matmult} and \ref{sec:conv}, we elucidate the most important technical contributions of the paper, namely the  \textit{Gazelle Linear Algebra Kernels} for fast matrix-vector multiplication and convolution. We then present detailed benchmarks on the implementation of the \textit{Gazelle Homomorphic Layer} and the linear algebra kernels in Section~\ref{sec:impl}. Finally, we describe the evaluation of neural networks such as ones trained on the MNIST or CIFAR-10 datasets and compare \gazelle's performance to prior work in Section~\ref{sec:networks}.

\section{Packed Additively Homomorphic Encryption}
\label{sec:prelimsHE}

In this section, we describe a clean abstraction of packed additively homomorphic encryption ($\pahename$) schemes that we will use through the rest of the paper. As suggested by the name, the abstraction will support packing multiple plaintexts into a single ciphertext, performing SIMD homomorphic additions ($\SIMDadd$) and scalar multiplications ($\SIMDmult$), and permuting the plaintext slots ($\Perm$). In particular, we will never need or use homomorphic multiplication of two ciphertexts. This abstraction can be instantiated with essentially all modern lattice-based homomorphic encryption schemes, e.g.,~\cite{BGV12,GHS12,B12,FV12}.

For the purposes of this paper, a private-key $\pahename$ suffices. In such an encryption scheme, we have a (randomized) encryption algorithm ($\paheenc$) that takes a plaintext message vector $\vecu$ from some message space and encrypts it using a key $\textsf{sk}$ into a ciphertext denoted as $[\vecu]$, and a (deterministic) decryption algorithm ($\pahedec$) that takes the ciphertext $[\vecu]$ and the key $\textsf{sk}$ and recovers the message $\vecu$. Finally, we also have a (randomized) homomorphic evaluation algorithm ($\paheeval$) that takes as input one or more ciphertexts that encrypt messages $M_0, M_1,\ldots$, and outputs another ciphertext that encrypts a message $M = f(M_0,M_1,\ldots)$ for some function $f$ constructed using the $\SIMDadd$, $\SIMDmult$ and $\Perm$ operations.

We require two security properties from a homomorphic encryption scheme: (1) IND-CPA Security, which requires that ciphertexts of any two messages $\vecu$ and $\vecu'$ are computationally indistinguishable; and (2) Function Privacy, which requires that the ciphertext generated by homomorphic evaluation, together with the private key $\textsf{sk}$, reveals the underlying message, namely the output $f(\cdot)$, but does not reveal any other information about the function $f$.

The lattice-based $\pahename$ constructions that we consider in this paper are parameterized by four constants: (1) the cyclotomic order $m$, (2) the ciphertext modulus $q$, (3) the plaintext modulus $p$ and (4) the standard deviation $\sigma$ of a symmetric discrete Gaussian noise distribution ($\chi$). 

The number of slots in a packed $\pahename$ ciphertext is given by $n = \phi(m)$ where $\phi$ is the Euler Totient function. Thus, plaintexts can be viewed as length-$n$ vectors over $\mathbb{Z}_p$ and ciphertexts are viewed as length-$n$ vectors over $\mathbb{Z}_q$. All fresh ciphertexts start with an inherent noise $\eta$ sampled from the noise distribution $\chi$. As homomorphic computations are performed $\eta$ grows continually. Correctness of $\pahedec$ is predicated on the fact that $|\eta| < q/(2p)$, thus setting an upper bound on the complexity of the possible computations.

In order to guarantee security we require a minimum value of $\sigma$ (based on $q$ and $n$), $q \equiv 1 \mod m$ and $p$ is co-prime to $q$. Additionally, in order to minimize noise growth in the homomorphic operations we require that the magnitude of $r \equiv q \mod p$ be as small as possible. This when combined with the security constraint results in an optimal value of $r = \pm1$. 

In the sequel, we describe in detail the three basic operations supported by the homomorphic encryption schemes together with their associated asymptotic cost in terms of (a) the run-time, and (b) the noise growth. Later, in Section~\ref{sec:impl}, we will provide concrete micro-benchmarks for each of these operations implemented in the $\mbox{\sc gazelle}$ library.

\subsection{Ciphertext Addition: $\SIMDadd$}
Given ciphertexts $[\vecu]$ and $[\vecv]$, $\SIMDadd$ outputs an encryption of their componentwise sum, namely $[\vecu + \vecv]$. 
    
The asymptotic run-time for homomorphic addition is $n\cdot \CostAdd(q)$, where $\CostAdd(q)$ is the run-time for adding two numbers in $\mathbb{Z}_q = \{0,1,\ldots,q-1\}$. The noise growth is at most $\eta_{\vecu} + \eta_{\vecv}$ where $\eta_\vecu$ (resp. $\eta_\vecv$) is the amount of noise in $[\vecu]$ (resp. in $[\vecv]$).

\subsection{Scalar Multiplication: $\SIMDmult$}
If the plaintext modulus is chosen such that $p \equiv 1 \mod m$, we can also support a SIMD compenentwise product. Thus given a ciphertext $[\vecu]$ and a plaintext $\vecv$, we can output an encryption $[\vecu \circ \vecv]$ (where $\circ$ denotes component-wise multiplication of vectors).
    
The asymptotic run-time for homomorphic scalar multiplication is $n \cdot \CostMult(q)$, where $\CostMult(q)$ is the run-time for multiplying two numbers in $\mathbb{Z}_q$. The noise growth is at most $\overmlt \cdot \eta_{\vecu}$ where $\overmlt \approx ||\vecv||^\prime_{\infty} \cdot \sqrt{n}$ is the {\em multiplicative noise growth} of the SIMD scalar multiplication operation.

For a reader familiar with homomorphic encryption schemes, we note that $||\vecv||^\prime_{\infty}$ is the largest value in the {\em coefficient representation} of the packed plaintext vector $\vecv$, and thus, even a binary plaintext vector can result in $\overmlt$ as high as $p \cdot \sqrt{n}$. In practice, we alleviate this large multiplicative noise growth by bit-decomposing the coefficient representation of $\vecv$ into $\log(p/2^{\wpt})$ many $\wpt$-sized chunks $\vecv_k$ such that $\vecv = \sum 2^{\wpt \cdot k}\cdot\vecv_k$. We refer to $\wpt$ as the plaintext window size. 

We can now represent the product $[\vecu \circ \vecv]$ as $\sum [\vecu_k \circ \vecv_k]$ where $\vecu_k = [2^{\wpt \cdot k} \cdot \vecu]$. Since $||\vecv_k||^\prime_{\infty} \leq 2^{\wpt}$ the total noise in the multiplication is bounded by $\sum_k 2^{\wpt} \cdot \sqrt{n} \cdot \eta_{\vecu_k}$ as opposed to $p \cdot \sqrt{n} \cdot \eta_{\vecu}$. The only caveat is that we need access to low noise encryptions $[\vecu_k]$ as opposed to just $[\vecu]$ as in the direct approach.

\subsection{Scalar Multiplication: $\Perm$}
\begin{figure}
    \centering
    \includegraphics[scale=0.4]{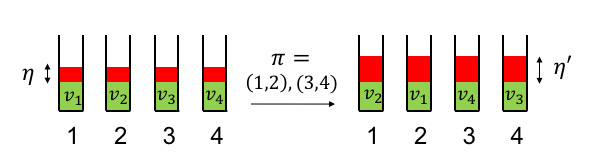}
    \caption{A Plaintext Permutation in action. The permutation $\pi$ in this example swaps the first and the second slots, and also the third and fourth slots. The operation incurs a noise growth from $\eta$ to $\eta' \approx \eta + \etarot$. Here, $\etarot \approx n\log q\cdot \eta_0$ where $\eta_0$ is some small ``base noise''.}
    \label{fig:rot}
\end{figure}

Given a ciphertext $[\vecu]$ and one of a set of {\em primitive permutations} $\pi$ defined by the scheme, the $\Perm$ operation outputs a ciphertext $[\vecu_{\pi}]$, where   $\vecu_\pi$ is defined as $(u_{\pi(1)},u_{\pi(2)},\ldots, u_{\pi(n)})$, namely the vector $\vecu$ whose slots are permuted according to the permutation $\pi$. The set of permutations that can be supported depends on the structure of the multiplicative group $\mod m$ i.e. $(\mathbb{Z}/m\mathbb{Z})^{\times}$. When $m$ is prime, we have $n$ ($= m-1$) slots and the permutation group supports all cyclic rotations of the slots, i.e. it is isomorphic to $C_{n}$ (the cyclic group of order $n$). When $m$ is a sufficiently large power of two $(m = 2^k, m \geq 8)$, we have $n = 2^{k-1}$ and the set of permutations is isomorphic to the set of half-rotations i.e. $C_{n/2}\times C_2$, as illustrated in Figure~\ref{fig:rot}.

Permutations are by far the most expensive operations in a homomorphic encryption scheme. A single permutation costs as much as performing a number theoretic transform ($\NTT$), the analog of the discrete Fourier transform, plus the cost of $\Theta(\log q)$ inverse number theoretic transforms ($\NTT^{-1}$). Since $\NTT$ and $\NTT^{-1}$ have an asymptotic cost of $\Theta(n\log n)$, the cost is therefore $\Theta(n\log n \log q)$. The noise growth is additive, namely, $\eta_{\vecu_\pi} = \eta_{\vecu}+ \etarot$ where $\etarot$ is the {\em additive noise growth} of a permutation operation.

\begin{figure}
    \centering
    \includegraphics[scale=0.4]{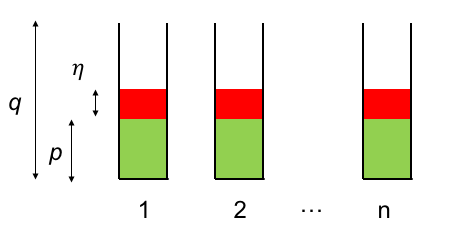}
    \caption{Ciphertext Structure and Operations. Here, $n$ is the number of slots, $q$ is the size of ciphertext space (so a ciphertext required $\lceil \log_2 q \rceil$ bits to represent), $p$ is the size of the plaintext space (so a plaintext can have at most $\lfloor \log_2 p \rfloor$ bits), and $\eta$ is the amount of noise in the ciphertext.}
    \label{fig:slots}
\end{figure}

\subsection{Paillier vs. Lattice-based $\pahename$}

The $\pahename$ scheme used in \gazelle\ is dramatically more efficient than conventional Paillier based AHE. Homomorphic addition of two Paillier ciphertexts corresponds to a modular multiplication modulo a large RSA-like modulus (~2048bits) as opposed to a simple addition $\mod q$ as seen in $\SIMDadd$. Similarly multiplication by a plaintext turns into a modular exponentiation for Paillier. Furthermore the large sizes of the Paillier ciphertexts makes encryption of single small integers extremely bandwidth-inefficient. In contrast, the notion of packing provided by lattice-based schemes provides us with a SIMD way of packing many integers into one ciphertext, as well as SIMD evaluation algorithms. We are aware of one system~\cite{SSW09} that tries to use Paillier in a SIMD fashion; however, this lacks two crucial components of lattice-based AHE, namely the facility to multiply each slot with a separate scalar, and the facility to permute the slots. We are also aware of a method of mitigating the first of these shortcomings~\cite{IW06}, but not the second. Our fast homomorphic implementation of linear algebra uses both these features of lattice-based AHE, making Paillier an inherently unsuitable substitute.

\subsection{Parameter Selection for $\pahename$}

Parameter selection for $\pahename$ requires a delicate balance between the homomorphic evaluation capabilities and the target security level. We detail our procedure for parameter selection to meet a target security level of 128 bits. We first set our plaintext modulus to be ~20 bits to represent the fixed point inputs (the bit-length of each pixel in an image) and partial sums generated during the neural network evaluation. Next, we require that the ciphertext modulus be close to, but less than, 64 bits in order to ensure that each ciphertext slot fits in a single machine word while maximizing the potential noise margin available during homomorphic computation. 

The $\Perm$ operation in particular presents an interesting tradeoff between the simplicity of possible rotations and the computational efficiency of the number-theoretic transform (NTT). A prime $m$ results in a (simpler) cyclic permutation group but necessitates the use of an expensive Bluestein transform. Conversely, the use of $m = 2^k$ allows for a $~8\times$ more efficient Cooley-Tukey style NTT at the cost of an awkward permutation group that only allows half-rotations. In this work, we opt for the latter and adapt our linear algebra kernels to deal with the structure of the permutation group. Based on the analysis of \cite{albrecht2015concrete}, we set $m = 4096$ and $\sigma = 4$ to obtain our desired security level.

Our chosen bit-width for $q$, namely $60$ bits, allows for lazy reduction, i.e. multiple additions may be performed without overflowing a machine word before a reduction is necessary. Additionally, even when $q$ is close to the machine word-size, we can replace modular reduction with a simple sequence of addition, subtraction and multiplications. This is done by choosing $q$ to be a pseudo-Mersenne number.

Next, we detail a technique to generate prime moduli that satisfy the above correctness and efficiency properties, namely:
\begin{enumerate}
    \item $q \equiv 1 \pmod m$
    \item $p \equiv 1 \pmod m$
    \item $|q \bmod p| = |r| \approx 1$
    \item $q$ is pseudo-Mersenne, i.e. $q = 2^{60} - \delta, (\delta < \sqrt{q})$
\end{enumerate}
Below, we describe a fast method to generate $p$ and $q$ (We remark that the obvious way to do this requires at least $p\approx 2^{20}$ primality tests, even to satisfy the first three conditions).

Since we have chosen $m$ to be a power of two, we observe that $\delta \equiv -1 \pmod m$. Moverover $r \equiv q \pmod p$ implies that $\delta \equiv (q-r) \pmod p$. These two CRT expressions for $\delta$ imply that given a prime $p$ and residue $r$, there exists a unique minimal value of $\delta \mod (p\cdot m)$. 

Based on this insight our prime selection procedure can be broken down into three steps:
\begin{enumerate}
    \item Sample for $p \equiv 1 \mod m$ and sieve the prime candidates.
    \item For each candidate $p$, compute the potential $2|r|$ candidates for $\delta$ (and thus $q$). 
    \item If $q$ is prime and $\delta$ is sufficiently small accept the pair $(p, q)$.
\end{enumerate}

Heuristically, this procedure needs $\log(q)(p\cdot m)/(2|r|\sqrt{q})$ candidate primes $p$ to sieve out a suitable $q$. Since $p\approx 2^{20}$ and $q \approx 2^{64}$ in our setting, this procedure is very fast. A list of reduction-friendly primes generated by this approach is tabulated in Table~\ref{tab:primes}. Finally note that when $\lfloor\log(p)\rfloor\cdot3 < 64$ we can use Barrett reduction to speed-up reduction $\bmod p$.

\begin{table}
\centering
    \caption{Prime Selection for $\pahename$}
\begin{tabular}{c c c c}
\toprule
$\lfloor\log(p)\rfloor$ & $p$ & $q$ & $|r|$ \\ \midrule
$18$ & $307201$ & $2^{60}-2^{12}\cdot 63548 + 1$ & 1 \\
$22$ & $5324801$ & $2^{60}-2^{12}\cdot 122130 + 1$ & 1 \\
$26$ & $115351553$ & $2^{60}-2^{12}\cdot 9259 + 1$ & 1 \\
$30$ & $1316638721$ & $2^{60}-2^{12}\cdot 54778 + 1$ & 2 \\ \bottomrule
\end{tabular}
\label{tab:primes}
\end{table}

The impact of the selection of reduction-friendly primes on the performance of the $\pahename$ scheme is described in section \ref{sec:impl}.

\section{Our Protocol at a High Level}
\label{sec:protocol}

\begin{figure}
    \centering
    \includegraphics[scale=0.35]{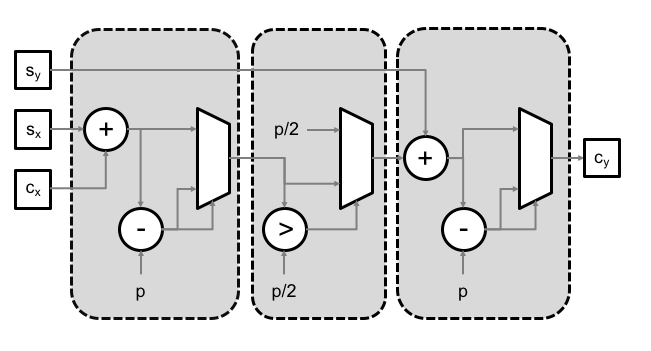}
    \caption{Our optimized circuit for step (b) namely Yao garbling. The ``+'' gates refer to an integer addition circuit and ``-'' refers to an integer subtraction circuit. The trapeziods are multiplexers and the ``$>$'' refers to the circuit that outputs $1$ if and only if the input is larger than $p/2$.}
    \label{fig:yaockt}
\end{figure}

Our protocol for solving the above problem is based on the alternating use of packed additively homomorphic encryption ($\pahename$) and garbled circuits (GC) to evaluate the neural network under consideration. Thus, the client $B$ first encrypts their input using the \gazelle\ SIMD linear homomorphic encryption scheme and sends it to the server $A$. The server $A$ first uses the \gazelle\ homomorphic neural network kernel for the first layer (which is either convolution or fully connected). The result is a packed ciphertext that contains the input to the first non-linear (ReLU) layer. 

To evaluate the first non-linear layer, we employ a garbled circuit based evaluation protocol. Our starting point is the scenario where $A$ holds a ciphertext $[\vecx]$ (where $\vecx$ is a vector) and $B$ holds the private key. $A$ and $B$ together do the following: 
\begin{description}
    \item[(a)] {\em Translate from Ciphertext to Shares:} The first step is to convert this into the scenario where $A$ and $B$ hold an additive secret sharing of $\vecx$. This is accomplished by the server $A$ adding a random vector $\vecr$ to her ciphertext homomorphically to obtain an encryption $[\vecx + \vecr]$ and sends it to the client $B$. The client $B$ decrypts it; the server $A$ sets her share $s_x = \vecr$ and $B$ sets his share $c_x = \vecx+ \vecr \pmod{p}$. This is clearly an additive (arithmetic) secret sharing of $\vecx$.
    \item[(b)] {\em Yao Garbled Circuit Evaluation:} We now wish to run the Yao garbled circuit protocol for the non-linear activation functions $f$ (in parallel for each component of $\vecx$) to get a secret sharing of the output $\vecy = f(\vecx)$. This is done using our circuit from Figure~\ref{fig:yaockt}, described in more detail below. The output of the garbled circuit evaluation is a pair of shares $s_y$ (for the server) and $c_y$ (for the client) such that $s_y + c_y = y \pmod{p}$.
    \item[(c)] {\em Translate back from Shares to a Ciphertext}: The client $A$ encrypts her share $c_y$ using the homomorphic encryption scheme and sends it to $B$; $B$ in turn homomorphically adds his share $s_y$ to obtain an encryption of $c_y + s_y = y = f(x)$.
\end{description}

Once this is done, we are back where we started. The next linear layer (either fully connected or convolutional) is evaluated using the \gazelle\ homomorphic neural network kernel, followed by Yao's garbled circuit protocol for the next non-linear layer, so we rinse and repeat until we evaluate the full network. We make the following two observations about our proposed protocols:
\begin{enumerate}
    \item By using AHE for the linear layers, we ensure that the communication complexity of protocol is linear in the number of layers and the size of inputs for each layer.
    \item At the end of the garbled circuit protocol we have an additive share that can be encrypted afresh. As such, we can view the re-encryption as an interactive bootstrapping procedure that clears the noise introduced by any previous homomorphic operation.
\end{enumerate}

For the second step of the outline above, we employ the {\em Boolean} circuit described in Figure~\ref{fig:yaockt}.
The circuit takes as input three vectors: $s_x = \vecr$ and $s_y = \vecr'$ (chosen at random) from the server, and $c_x$ from the client. The first block of the circuit computes the arithmetic sum of $s_x$ and $c_x$ over the integers and subtracts $p$ from to obtain the result mod $p$. (The decision of whether to subtract $p$ or not is made by the multiplexer). The second block of the circuit computes a ReLU function. The third block adds the result to $s_y$ to obtain the client's share of $y$, namely $c_y$. For more detailed benchmarks on the ReLU and MaxPool garbled circuit implementations, we refer the reader to Section~\ref{sec:networks}.

In our evaluations, we consider ReLU, Max-Pool and the square activation functions, the first two are by far the most commonly used ones in convolutional neural network design~\cite{KSH12,googlenet,vggnet,resnet}. Note that the square activation function popularized for secure neural network evaluation in \cite{cryptonets} can be efficiently implemented by a simple interactive protocol that use the $\pahename$ scheme to generate the cross-terms.

\begin{table*}
\centering
\begin{threeparttable}
    \caption[Comparison of matrix-vector product algorithms]{Comparing matrix-vector product algorithms by operation count, noise growth and number of output ciphertexts}
    
    \begin{tabular}{ccccccc} \toprule
    & $\Perm$ (Hoisted)\tnote{a} & $\Perm$ & $\SIMDmult$ & $\SIMDadd$ & Noise & $\mathsf{\# out\_ct}$\tnote{b} \\
    \midrule
    
    \multirow{2}{*}{Na\"{i}ve} & \multirow{2}{*}{$0$} & \multirow{2}{*}{$n_o \cdot \log n_i$} & \multirow{2}{*}{$n_o$} & \multirow{2}{*}{$n_o\cdot \log n_i$} & $\eta_{\mathsf{naive}} := \eta_0 \cdot \overmlt\cdot n_i$ & \multirow{2}{*}{$n_o$} \\
    & & & & & $+ \etarot \cdot (n_i-1)$ & \\
    \midrule
    
    Na\"{i}ve & \multirow{2}{*}{$0$} & \multirow{2}{*}{$n_o \cdot \log n_i + n_o -1$}  & \multirow{2}{*}{$2\cdot n_o$} & \multirow{2}{*}{$n_o\cdot \log n_i + n_o$}  & $\eta_{\mathsf{naive}} \cdot \overmlt \cdot n_o$  & \multirow{2}{*}{$1$} \\
    (Output packed) &  &  &  &  & $+\etarot\cdot (n_o-1)$ & \\
    \midrule
    
    Na\"{i}ve & \multirow{2}{*}{$0$} & \multirow{2}{*}{$\frac{n_o \cdot n_i}{n} \cdot \log n_i $} & \multirow{2}{*}{$\frac{n_o \cdot n_i}{n}$} & \multirow{2}{*}{$\frac{n_o \cdot n_i}{n} \cdot \log n_i$} & $\eta_0 \cdot \overmlt\cdot n_i$ & \multirow{2}{*}{$\frac{n_o \cdot n_i}{n}$} \\
    (Input packed) & & & & & $+ \etarot \cdot (n_i-1)$ & \\
    \midrule
    
    Diagonal & $n_i-1$ & $0$ & $n_i$ & $n_i$& $(\eta_0+\etarot)\cdot \overmlt \cdot n_i$ & $1$ \\
    \midrule
    
    \multirow{2}{*}{Hybrid} & \multirow{2}{*}{$\frac{n_o \cdot n_i}{n} -1$} & \multirow{2}{*}{$\log \frac{n}{n_o}$} & \multirow{2}{*}{$\frac{n_o \cdot n_i}{n}$} & \multirow{2}{*}{$\frac{n_o \cdot n_i}{n} + \log \frac{n}{n_o}$}  & $ (\eta_0+\etarot)\cdot \overmlt \cdot n_i$& \multirow{2}{*}{$1$} \\
    & & & & & $+\etarot\cdot (\frac{n_i}{n_o} - 1)$ & \\
    \bottomrule
    \end{tabular}
    
    \begin{tablenotes}
    \footnotesize
    \item[a] Rotations of the input with a common $\PermDecomp$
    \item[b] Number of output ciphertexts\\
    \item[c] All logarithms are to base $2$
    \end{tablenotes}
\end{threeparttable}
\label{tab:matmult}
\end{table*}

\section{Fast Homomorphic Matrix-Vector Multiplication}
\label{sec:matmult}

We next describe the \gazelle\  homomorphic linear algebra kernels that compute matrix-vector products (for $\fc$ layers) and 2-d convolutions (for $\conv$ layers). In this section, we focus on matrix-vector product kernels which multiply a plaintext matrix with an encrypted vector. We start with the easiest to explain (but the slowest and most communication-inefficient) methods and move on to describing optimizations that make matrix-vector multiplication much faster. In particular, our {\bf hybrid method} (see Table~\ref{tab:matmult} and the description below) gives us the best performance among all our homomorphic matrix-vector multiplication methods. For example, multiplying a $128\times 1024$ matrix with a length-$1024$ vector using our hybrid scheme takes about $16$ms on a commodity machine. (For detailed benchmarks, we refer the reader to Section~\ref{subsec:micro}). In all the subsequent examples, we will use an $\fc$ layer with $n_i$ inputs and $n_o$ outputs as a running example. For simplicity of presentation, unless stated otherwise we assume that $n$, $n_i$ and $n_o$ are powers of two. Similarly we assume that $n_o$ and $n_i$ are smaller than $n$. If not, we can split the original matrix into $n \times n$ sized blocks that are processed independently. 


\subsection{The Na\"{i}ve Method}

\begin{figure}
    \centering
    \includegraphics[scale=0.85]{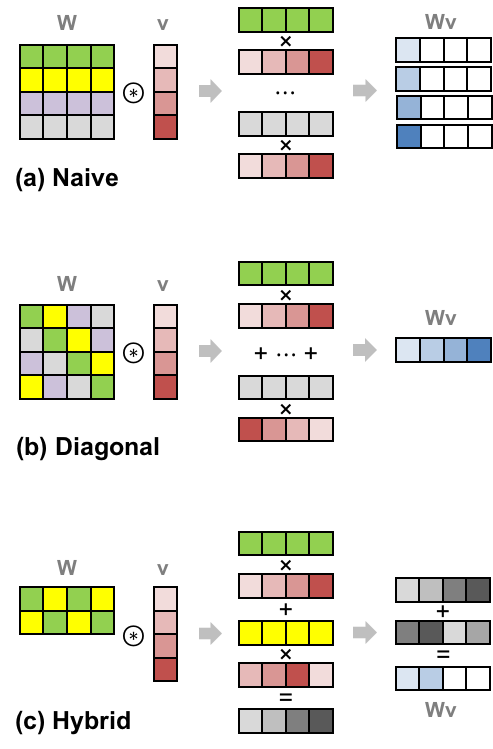}
    \caption{The na\"{i}ve method is illustrated on the left and the diagonal method of Halevi and Shoup~\cite{HS14} is illustrated on the right. The entries in a single color live in the same ciphertext. The key feature of the diagonal method is that no two elements of the matrix that influence the same output element appear with the same color.}
    \label{fig:matmult}
\end{figure}

In the na\"{i}ve method, each row of the $n_o\times n_i$ plaintext weight matrix $\matW$ is encoded into a separate plaintext vectors (see Figure~\ref{fig:matmult}). Each such vector is of length $n$; where the first $n_i$ entries contain the corresponding row of the matrix and the other entries are padded with $0$. These plaintext vectors are denoted $\vecw_0,\vecw_1,\ldots,\vecw_{(n_o-1)}$. We then use $\SIMDmult$ to compute the componentwise product of with the encrypted input vector $[\vecv]$ to get $[\vecu_i] = [\vecw_i \circ \vecv]$. In order to compute the inner-product what we need is actually the sum of the entries in each of these vectors $\vecu_i$. 

This can be achieved by a ``rotate-and-sum'' algorithm, where we first rotate the entries of  $[\vecu_i]$ by $n_i/2$ positions. The result is a ciphertext whose first $n_i/2$ entries contain the sum of the first and second halves of $\vecu_i$. One can then repeat this process for $\log_2 n_i$ iterations, rotating by half the previous rotation on each iteration, to get a ciphertext whose first slot contains the first component of $\matW\vecv$. By repeating this procedure for each of the $n_o$ rows we get $n_o$ ciphertexts, each containing one element of the result. 

Based on this description, we can derive the following performance characteristics for the na\"{i}ve method:

\begin{itemize}
    \item The total cost is $n_o$ SIMD scalar multiplications, $n_o\cdot \log_2 n$ rotations (automorphisms) and $n_o\cdot \log_2 n$ SIMD additions. 
    \item The noise grows from $\eta$ to $\eta \cdot \overmlt \cdot n + \etarot\cdot (n-1)$ where $\overmlt$ is the multiplicative noise growth factor for SIMD multiplication and $\etarot$ is the additive noise growth for a rotation. This is because the one SIMD multiplication turns the noise from $\eta \mapsto \eta\cdot \overmlt$, and the sequence of rotations and additions grows the noise as follows:
    \[ \eta\cdot \overmlt \mapsto (\eta\cdot \overmlt)\cdot 2 + \etarot \mapsto (\eta\cdot \overmlt)\cdot 4 + \etarot\cdot 3 \mapsto \ldots \]
    which gives us the above result.
    \item Finally, this process produces $n_o$ many ciphertexts each one containing just one component of the result. 
\end{itemize}

This last fact turns out to be an unacceptable efficiency barrier. In particular, the total network bandwidth becomes quadratic in the input size and thus contradicts the entire rationale of using $\pahename$ for linear algebra. Ideally, we want the entire result to come out in packed form {\em in a single ciphertext} (assuming, of course, that $n_o \leq n$).

A final subtle point that needs to noted is that if $n$ is not a power of two, then we can continue to use the same rotations as before, but all slots except the first slot leak information about partial sums. We therefore \textit{must} add a random number to these slots to destroy this extraneous information about the partial sums.

\subsection{Output Packing} 

The very first thought to mitigate the ciphertext blowup issue we just encountered is to take the many output ciphertexts and somehow pack the results into one. Indeed, this can be done by (a) doing a SIMD scalar multiplication which zeroes out all but the first coordinate of each of the $\out$ ciphertexts; (b) rotating each of them by the appropriate amount so that the numbers are lined up in different slots; and (c) adding all of them together.

Unfortunately, this results in unacceptable noise growth. The underlying reason is that we need to perform two serial SIMD scalar multiplications (resulting in an $\overmlt^2$ factor; see Table~\ref{tab:matmult}). For most practical settings, this noise growth forces us to use ciphertext moduli that are larger $64$ bits, thus overflowing the machine word. This necessitates the use of a Double Chinese Remainder Theorem (DCRT) representation similar to \cite{GHS12} which substantially slows down computation. Instead we use an algorithmic approach to control noise growth allowing the use of smaller moduli and avoiding the need for DCRT.

\subsection{Input Packing} 

Before moving on to more complex techniques we describe an orthogonal approach to improve the na\"{i}ve method when $n_i \ll n$. The idea is to pack multiple copies of the input into a single ciphertext. This allows us better utilization of the slots by computing multiple outputs in parallel. 

In detail we can (a) pack $n/n_i$ many different rows into a single plaintext vector; (b) pack $n/n_i$ {\em copies} of the input vector into a single ciphertext; and (c) perform the rest of the na\"{i}ve method as-is except that the rotations are not applied to the whole ciphertext but block-by-block (thus requiring $\log(n_i)$ many rotations). Roughly speaking, this achieves communication and computation as if the number of rows of the matrix were $n_o' = {(n_o\times n_i)}/{n}$ instead of $n_o$. When $n_i \ll n$, we have $n_o' \ll n_o$. 

\subsection{The Diagonal Method}

The diagonal method as described in the work of Halevi and Shoup~\cite{HS14} (and implemented in \cite{HELib}) provides another potential solution to the problem of a large number of output ciphertexts. The key high-level idea is to arrange the matrix elements in such a way that after the SIMD scalar multiplications, ``interacting elements'' of the matrix-vector product never appear in a single ciphertext. Here, ``interacting elements'' are the numbers that need to be added together to obtain the final result. The rationale is that if this happens, we never need to add two numbers that live in different slots of the same ciphertexts, thus avoiding ciphertext rotation.

To do this, we encode the diagonal of the matrix into a vector which is then SIMD scalar multiplied with the input vector. The second diagonal (namely, the elements $\matW_{0, 1}, \matW_{1,2}, \ldots, \matW_{n_o-1, 0}$) is encoded into another vector which is then SIMD scalar multiplied with a rotation (by one) of the input vector, and so on. Finally, all these vectors are added together to obtain the output vector {\em in one shot}.

The cost of the diagonal method is:
\begin{itemize}
    \item The total cost is $n_i$ SIMD scalar multiplications, $n_i-1$ rotations (automorphisms),  and $n_i-1$ SIMD additions. 
    \item The noise grows from $\eta$ to $(\eta+\etarot)\cdot \overmlt \times n_i$ which, for the parameters we use, is larger than that of the na\"{i}ve method, but much better than the na\"{i}ve method with output packing. Roughly speaking, the reason is that in the diagonal method, since rotations are performed before scalar multiplication, the noise growth has a $\etarot\cdot \overmlt$ factor whereas in the na\"{i}ve method, the order is reversed resulting in a $\overmlt + \etarot$ factor.
    \item Finally, this process produces {\em a single} ciphertext that has the entire output vector in packed form already. 
\end{itemize}

In our setting (and we believe in most reasonable settings), the additional noise growth is an acceptable compromise given the large gain in the output length and the corresponding gain in the bandwidth and the overall run-time. Furthermore, the fact that all rotations happen on the input ciphertexts prove to be very important for an optimization of \cite{HS17} we describe below, called ``hoisting'', which lets us amortize the cost of many {\em input} rotations.

\subsection{Book-keeping: Hoisting}

The hoisting optimization reduces the cost of the ciphertext rotation when the \textit{same} ciphertext must be rotated by multiple shift amounts. The idea, roughly speaking, is to ``look inside'' the ciphertext rotation operation, and hoist out the part of the computation that would be common to these rotations and then compute it only once thus amortizing it over many rotations. It turns out that this common computation involves computing the $\NTT^{-1}$ (taking the ciphertext to the coefficient domain), followed by a $\wrelin$-bit decomposition that splits the ciphertext $\lceil (\log_2 q)/\wrelin\rceil$ ciphertexts and finally takes these ciphertexts back to the evaluation domain using separate applications of $\NTT$. The parameter $\wrelin$ is called the relinearization window and represents a tradeoff between the speed and noise growth of the $\Perm$ operation. This computation, which we denote as $\PermDecomp$, has $\Theta\left(n \log n\right)$ complexity because of the number theoretic transforms. In contrast, the independent computation in each rotation, denoted by $\PermAuto$, is a simple $\Theta\left(n\right)$ multiply and accumulate operation. As such, hoisting can provide substantial savings in contrast with direct applications of the $\Perm$ operation and this is also borne out by the benchmarks in Table~\ref{tab:micro-perm}.

\subsection{A Hybrid Approach}

One issue with the diagonal approach is that the number of $\Perm$ is equal to $n_i$. In the context of $\fc$ layers $n_o$ is often much lower than $n_i$ and hence it is desirable to have a method where the $\Perm$ is close to $n_o$. Our hybrid scheme achieves this by combining the best aspects of the na\"{i}ve and diagonal schemes. We first extended the idea of diagonals for a square matrix to squat rectangular weight matrices as shown in Figure~\ref{fig:matmult} and then pack the weights along these extended diagonals into plaintext vectors. These plaintext vectors are then multiplied with $n_o$ rotations of the input ciphertext similar to the diagonal method. Once this is done we are left with a single ciphertext that contains $n/n_o$ chunks each contains a partial sum of the $n_o$ outputs. We can proceed similar to the na\"{i}ve method to accumulate these using a ``rotate-and-sum'' algorithm. 

We implement an input packed variant of the hybrid method and the performance and noise growth characteristics (following a straightforward derivation) are described in Table~\ref{tab:matmult}. We note that hybrid method trades off hoistable input rotations in the Diagonal method for output rotations on distinct ciphertexts (which cannot be ``hoisted out''). However, the decrease in the number of input rotations is multiplicative while the corresponding increase in the number of output rotations is the logarithm of the same multiplicative factor. As such, the hybrid method almost always outperforms the Naive and Diagonal methods. We present detailed benchmarks over a selection of matrix sizes in Table~\ref{tab:micro-mm}.

\begin{figure}
    \centering
    \includegraphics[scale=0.6]{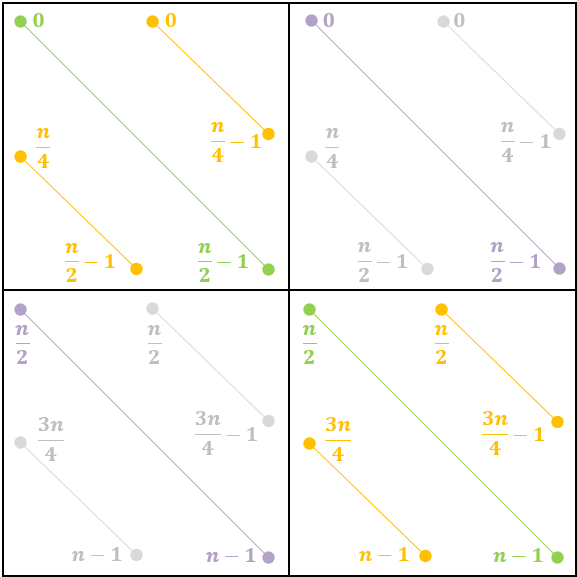}
    \caption{Four example extended digaonals after accounting for the rotation group structure}
    \label{fig:matmult-halfrot}
\end{figure}

We close this section with two implementation details. First, recall that in order to enable faster $\NTT$, our parameter selection requires $n$ to be a power of two. As a result the permutation group we have access to is the group of half rotations ($C_{n/2} \times C_2$), i.e. the possible permutations are compositions of rotations by up to $n/2$ for the two $n/2$-sized segments, and swapping the two segments. The packing and diagonal selection in the hybrid approach are modified to account for this by adapting the definition of the extended diagonal to be those entries of $\matW$ that would be multiplied by the corresponding entries of the ciphertext when the above $\Perm$ operations are performed as shown in Figure~\ref{fig:matmult-halfrot}. Finally, as described in section \ref{sec:prelimsHE} we control the noise growth in $\SIMDmult$ using plaintext windows for the weight matrix $\matW$.

\section{Fast Homomorphic Convolutions}
\label{sec:conv}

We now move on the implementation of homomorphic kernels for $\conv$ layers. Analogous to the description of $\fc$ layers we will start with simpler (and correspondingly less efficient) techniques before moving on to our final optimized implementation. In our setting, the server has access to a plaintext filter and it is then provided encrypted input images, which it must homomorphically convolve with its filter to produce encrypted output images. As a running example for this section we will consider a $(f_w, f_h, c_i, c_o)$-$\conv$ layer with the ``\textit{same}'' padding scheme, where the input is specified by the tuple $(w_i, h_i, c_i)$.  In order to better emphasize the key ideas, we will split our presentation into two parts: first we will describe the single input single output (SISO) case, i.e. ($c_i = 1, c_o = 1$) followed by the more general case where we have multiple input and output channels, a subset of which may fit within a single ciphertext.

\subsection{Padded SISO}

\begin{figure}
    \centering
    \includegraphics[scale=0.55]{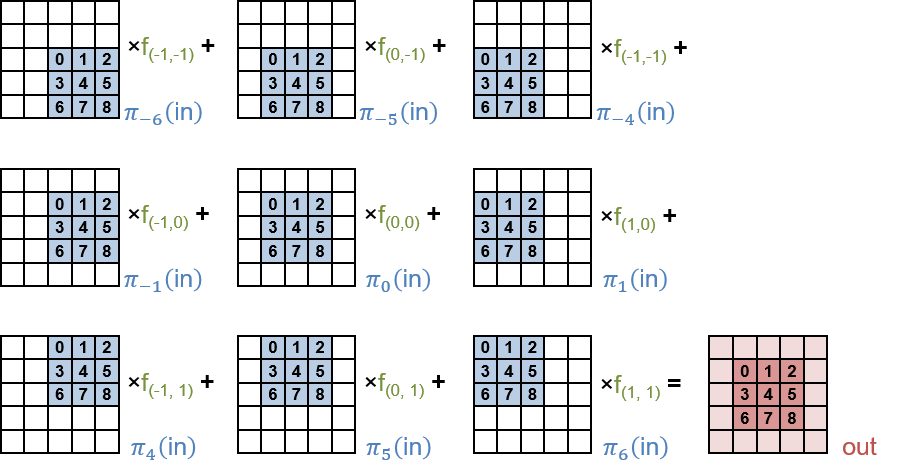}
    \caption{Padded SISO Convolution}
    \label{fig:conv-2d-padded}
\end{figure}

As seen in section \ref{sec:prelimsCNN}, \textit{same} style convolutions require that the input be zero-padded. As such, in this approach, we start with a zero-padded version of the input with $(f_w-1)/2$ zeros on the left and right edges and $(f_h-1)/2$ zeros on the top and bottom edges. We assume for now that this padded input image is small enough to fit within a single ciphertext i.e. $(w_i+f_w-1) \cdot (h_i+f_h-1) \leq n$ and is mapped to the ciphertext slots in a raster scan fashion. We then compute $f_w \cdot f_h$ rotations of the input and scale them by the corresponding filter coefficient as shown in Figure~\ref{fig:conv-2d-padded}. Since all the rotations are performed on a common input image, they can benefit from the hoisting optimization. Note that similar to the na\"{i}ve matrix-vector product algorithm, the values on the periphery of the output image leak partial products and must be obscured by adding random values.

\subsection{Packed SISO}

\begin{figure}
    \centering
    \includegraphics[scale=0.6]{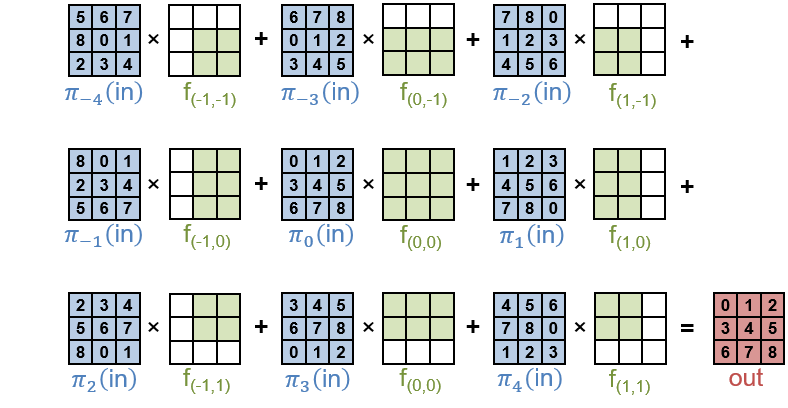}
    \caption{Packed SISO Convolution. (Zeros in the punctured plaintext shown in white.)}
    \label{fig:conv-2d-packed}
\end{figure}

While the above the technique computes the correct 2D-convolution it ends up wasting 
$(w_i + f_w - 1) \cdot (h_i +f_h - 1) - w_i\cdot h_i$ slots in zero padding. If either the input image is small or if the filter size is large, this can amount to a significant overhead. We resolve this issue by using the ability of our $\pahename$ scheme to multiply different slots with different scalars when performing $\SIMDmult$. As a result, we can pack the input tightly and generate $f_w \cdot f_h$ rotations. We then multiply these rotated ciphertexts with \textit{punctured plaintexts} which have zeros in the appropriate locations as shown in Figure~\ref{fig:conv-2d-packed}. Accumulating these products gives us a single ciphertext that, as a bonus feature, contains the convolution result without any leakage of partial information.

Finally, we note that the construction of the punctured plaintexts does not depend on either the encrypted image or the client key information and as such, the server can precompute these values once for multiple clients. We summarize these results in Table~\ref{tab:conv-siso}. 

\begin{table}
\centering
    \caption{Comparing SISO 2D-convolutions}
    \label{tab:conv-siso}
    \begin{tabular}{ccccc} \toprule
    & $\Perm$ & \#\ slots \\
    \midrule
    Padded & $f_w f_h-1$ & $(w_i+f_w-1)(h_i+f_h-1)$ \\
    Packed & $f_w f_h-1$ & $w_ih_i$ \\
    \bottomrule
    \end{tabular}
\end{table}

Now that we have seen how to compute a single 2D-convolution we will look at the more general multi-channel case.

\subsection{Single Channel per Ciphertext}

The straightforward approach for handling the multi-channel case is to encrypt the various channels into distinct ciphertexts. We can then SISO convolve these $c_i$-ciphertexts with each of the $c_o$ sets of filters to generate $c_o$ output ciphertexts. Note that although we need $c_o \cdot c_i\cdot f_h\cdot f_w$ $\SIMDadd$ and $\SIMDmult$ calls, just $c_i\cdot f_h\cdot f_w$ many $\Perm$ operations on the input suffice, since the rotated inputs can be reused to generate each of the $c_o$ outputs. Furthermore, each these rotation can be hoisted and hence we require just $c_i$ many $\PermDecomp$ calls and $c_i \cdot f_h\cdot f_w$ many $\PermAuto$ calls.

\subsection{Channel Packing}
Similar to input-packed matrix-vector products, the computation of multi-channel convolutions can be further sped up by packing multiple channels in a single ciphertext. We represent the number of channels that fit in a single ciphertext by $c_n$. Channel packing allows us to perform $c_n$-SISO convolutions in parallel in a SIMD fashion. We maximize this parallelism by using Packed SISO convolutions which enable us to tightly pack the input channels without the need for any additional padding.

For simplicity of presentation, we assume that both $c_i$ and $c_o$ are integral multiples of $c_n$. Our high level goal is to then start with $c_i/c_n$ input ciphertexts and end up with $c_o/c_n$ output ciphertexts where each of the input and output ciphertexts contains $c_n$ distinct channels. We achieve this in two steps: (a) convolve the input ciphertexts in a SISO fashion to generate $(c_o \cdot c_i)/c_n$ intermediate ciphertexts that contain all the $c_o\cdot c_i$-SISO convolutions and (b) accumulate these intermediate ciphertexts into output ciphertexts. 

\begin{figure}
    \centering
    \includegraphics[scale=0.4]{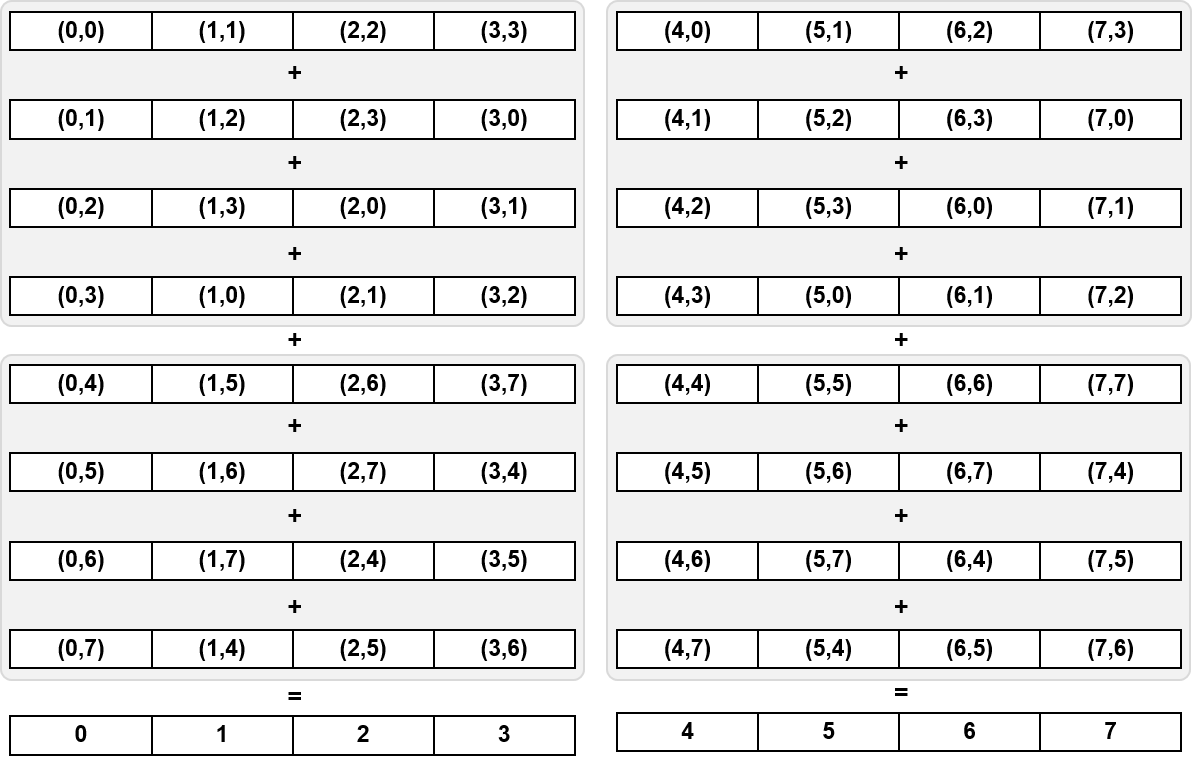}
    \caption{Diagonal Grouping for Intermediate Ciphertexts ($c_i = c_o = 8$ and $c_n = 4$)}
    \label{fig:conv-chn-pack}
\end{figure}

Since none of the input ciphertexts repeat an input channel, none of the intermediate ciphertexts can contain SISO convolutions corresponding to the same input channel. A similar constraint on the output ciphertexts implies that none of the intermediate ciphertexts contain SISO convolutions corresponding to the same output. In particular, a potential grouping of SISO convolutions that satisfies these constraints is the \textit{diagonal grouping}. More formally the $k^{th}$ intermediate ciphertext in the diagonal grouping contains the following ordered set of $c_n$-SISO convolutions: 
\begin{align*}
\{\  
    ( & \lfloor k/c_i \rfloor \cdot c_n + l, \\ & 
     \lfloor(k\bmod c_i)/c_n\rfloor \cdot c_n + ((k+l) \bmod c_n) 
    ) \mid \ l \in [0, c_n) 
\ \}
\end{align*}
where each tuple $(x_o, x_i)$ represents the SISO convolution corresponding to the output channel $x_o$ and input channel $x_i$. Given these intermediate ciphertexts, one can generate the output ciphertexts by simply accumulating the $c_o/c_n$-partitions of $c_i$ consecutive ciphertexts. We illustrate this grouping and accumulation when $c_i = c_o = 8$ and $c_n = 4$ in Figure~\ref{fig:conv-chn-pack}. Note that this grouping is very similar to {\em the diagonal style of computing matrix vector products}, with single slots now being replaced by entire SISO convolutions. 

Since the second step is just a simple accumulation of ciphertexts, the major computational complexity of the convolution arise in the computation of the intermediate ciphertexts. If we partition the set of intermediate ciphertexts into $c_n$-sized \textit{rotation sets} (shown in grey in Figure~\ref{fig:conv-chn-pack}), we see that each of the intermediate ciphertexts is generated by different rotations of the same input. This observation leads to two natural approaches to compute these intermediate ciphertexts.

\subsubsection*{Input Rotations}

In the first approach, we generate $c_n$ rotations of the every input ciphertext and then perform Packed SISO convolutions on each of these rotations to compute all the intermediate rotations required by $c_o/c_n$ rotation sets. Since each of the SISO convolutions requires $f_w \cdot f_h$ rotations, we require a total of $(c_n\cdot f_w\cdot f_h - 1)$ rotations (excluding the trivial rotation by zero) for each of the $c_i/c_n$ inputs. Finally we remark that by using the hoisting optimization we compute all these rotations by performing just $c_i/c_n$ $\PermDecomp$ operations.

\subsubsection*{Output Rotations}

The second approach is based on the realization that instead of generating $(c_n\cdot f_w\cdot f_h - 1)$ input rotations, we can reuse $(f_w\cdot f_h - 1)$ rotations in each rotation-set to generate $c_n$ convolutions and then simply rotate $(c_n - 1)$ of these to generate all the intermediate ciphertexts. This approach then reduces the number of input rotations by factor of $c_n$ while requiring $(c_n -1)$ for each of the $(c_i \cdot c_o)/c_n^2$ rotation sets. Note that while $(f_w\cdot f_h - 1)$ input rotations per input ciphertext can share a common $\PermDecomp$ each of the output rotations occur on a distinct ciphertext and cannot benefit from hoisting.

\begin{table*}
\centering
    \caption{Comparing multi-channel 2D-convolutions}
    \label{tab:conv-multi}
    \begin{tabular}{cccccc} \toprule
    & $\PermDecomp$ & $\Perm$ &
    $\mathsf{\# in\_ct}$ & $\mathsf{\# out\_ct}$ \\
    \midrule
    One Channel per CT & $c_i$ & $(f_w f_h-1)\cdot c_i$ & $c_i$ & $c_o$ \\
    \midrule
    Input Rotations & $\frac{c_i}{c_n}$ & $(c_n f_w f_h - 1)\cdot \frac{c_i}{c_n}$ & $\frac{c_i}{c_n}$ & $\frac{c_o}{c_n}$ \\
    \midrule
    Output Rotations& $\left(1+\frac{(c_n-1) \cdot c_o}{c_n}\right)\frac{c_i}{c_n}$ & $\left(f_w f_h - 1 + \frac{(c_n-1)\cdot c_o}{c_n}\right)\frac{c_i}{c_n}$ & $\frac{c_i}{c_n}$ & $\frac{c_o}{c_n}$ \\
    \bottomrule
    \end{tabular}
\end{table*}

We summarize these numbers in Table~\ref{tab:conv-multi}. The choice between the input and output rotation variants is an interesting trade-off that is governed by the size of the 2D filter. This trade-off is illustrated in more detail with concrete benchmarks in section \ref{sec:impl}. Finally, we remark that similar to the matrix-vector product computation, the convolution algorithms are also tweaked to work with the half-rotation permutation group and use plaintext windows to control the scalar multiplication noise growth.

\subsubsection*{Strided Convolutions} 
\begin{figure}
    \centering
    \includegraphics[scale=0.6]{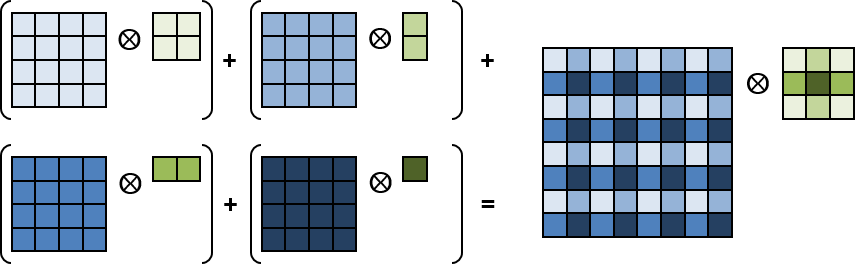}
    \caption{Decomposing a strided convolutions into simple convolutions ($f_w = f_h = 3$ and $s_x = s_y = 2$)}
    \label{fig:conv-stride}
\end{figure}

We handle strided convolutions by decomposing the strided convolution into a sum of simple convolutions each of which can be handled as above. We illustrate this case for $f_w = f_h = 3$ and $s_x = s_y = 2$ in Figure~\ref{fig:conv-stride}.
\vnote{Explain this in more detail in v2}

\subsubsection*{Low-noise Batched Convolutions}
We make one final remark on a potential application for padded SISO convolutions. Padded SISO convolutions are computed as a sum of rotated versions of the input images multiplied by corresponding constants $f_{x,y}$. The coefficient domain representation of these plaintext vectors is $(f_{x,y},0,\ldots,0)$. As a result, the noise growth factor is $\overmlt = f_{x,y}\cdot\sqrt{n}$ as opposed to $p\cdot\sqrt{n}$, consequently noise growth depends only on the value of the filter coefficients and {\em not} on the size of the plaintext space $p$. The direct use of this technique precludes the use of channel packing since the filter coefficients are channel dependent. One potential application that can mitigate this issue is when we want to classify a batch of multiple images. In this context, we can pack the same channel from multiple classifications allowing us to use a simple constant filter. This allows us to trade-off classification latency for higher throughput. Note however that similar to padded SISO convolutions, this has two problems: (a) it results in lower slot utilization compare to packed approaches, and (b) the padding scheme reveals the size of the filter.

\section{Implementation and Micro-benchmarks}
\label{sec:impl}

Next we describe the implementation of the \gazelle\ framework starting with the chosen cryptographic primitives (\ref{subsec:prim}). We then describe our evaluation test-bed (\ref{subsec:test-bed}) and finally conclude this section with detailed micro-benchmarks (\ref{subsec:micro}) for all the operations to highlight the individual contributions of the techniques described in the previous sections.

\subsection{Cryptographic Primitives}
\label{subsec:prim}

\gazelle\ needs two main cryptographic primitives for neural network inference: a packed additive homomorphic encryption ($\pahename$) scheme and a two-party secure computation (2PC) scheme. Parameters for both schemes are selected for a 128-bit security level. For the $\pahename$ scheme we instantiate the Brakerski-Fan-Vercauteren (BFV) scheme~\cite{B12,FV12}, which requires selection of the following parameters: ciphertext modulus ($q$), plaintext modulus ($p$), the number of SIMD slots ($n$) and the error parameter ($\sigma$). \vnote{how about the error?} Maximizing the $q/p$ ratio allows us to tolerate more noise, thus allowing for more computation. A plaintext modulus $p$ of 20 bits is enough to store all the intermediate values in the network computation \vnote{this sentence is a bit confusing}. This choice of the plaintext modulus size also allows for Barrett reduction on a 64-bit machine. The ciphertext modulus ($q$) is chosen to be a 60-bit psuedo-Mersenne prime that is slightly smaller than the native machine word on a 64-bit machine to enable lazy modular reductions. 

The selection of the number of slots is a more subtle trade-off between security and performance. In order to allow an efficient  implementation of the number-theoretic transform (NTT), the number of slots ($n$) must be a power of two. The amortized per-slot computational cost of both the $\SIMDadd$ and $\SIMDmult$ operations is $O(1)$, however the corresponding cost for the $\Perm$ operation is $O(\log n)$. This means that as $n$ increases, the computation becomes less efficient while on the other hand for a given $q$, a larger $n$ results in a higher security level. Hence we pick the smallest power of two that allows for a 128-bit security which in our case is $n=2048$.

For the 2PC framework, we use Yao's Garbled circuits~\cite{yao86}. The main reason for choosing Yao over Boolean secret sharing schemes (such as the Goldreich-Micali-Wigderson protocol~\cite{GMW87} and its derivatives) is that the constant number of rounds results in good performance over long latency links. Our garbling scheme is an extension of the one presented in JustGarble \cite{BHKR13} which we modify to also incorporate the Half-Gates optimization~\cite{ZRE15}. We base our oblivious transfer (OT) implementation on the classic Ishai-Kilian-Nissim-Petrank (IKNP) \cite{IKNP03} protocol from libOTe \cite{libote}. Since we use 2PC for implementing the ReLU, MaxPool and FHE-2PC transformation gadget, our circuit garbling phase only depends on the neural network topology and is independent of the client input. As such, we move it to the offline phase of the computation while the OT Extension and circuit evaluation is run during the online phase of the computation.


\subsection{Evaluation Setup}
\label{subsec:test-bed}

All benchmarks were generated using c4.xlarge AWS instances which provide a 4-threaded execution environment (on an Intel Xeon E5-2666 v3 2.90GHz CPU) with 7.5GB of system memory. Our experiments were conducted using Ubuntu 16.04.2 LTS (GNU/Linux 4.4.0-1041-aws) and our library was compiled using GCC 5.4.0 using the '-O3' optimization setting and enabling support for the AES-NI instruction set. Our schemes are evaluated in the LAN setting similar to previous work with both instances in the us-east-1a availability zone. 

\subsection{Micro-benchmarks}
\label{subsec:micro}

In order to isolate the impact of the various techniques and identify potential optimization opportunities, we first present micro-benchmarks for the individual operations. 

\subsubsection{Arithmetic and $\pahename$ Benchmarks}

We first benchmark the impact of the faster modular arithmetic on the NTT and the homomorphic evaluation run-times. Table \ref{tab:micro-ntt} shows that the use of a pseudo-Mersenne ciphertext modulus coupled with lazy modular reduction improves the NTT and inverse NTT by roughly $7\times$. Similarly Barrett reduction for the plaintext modulus improves the plaintext NTT runtimes by more than $5\times$. 
These run-time improvements are also reflected in the performance of the primitive homomorphic operations as shown in Table~\ref{tab:micro-he}.

\begin{table}
\centering
    \caption{Fast Reduction for NTT and Inv. NTT\vnote{say what cyc/bfly means.}}
    \label{tab:micro-ntt}
    \begin{tabular}{cccccc} \toprule
        \multirow{2}{*}{Operation} & 
            \multicolumn{2}{c}{Fast Reduction} & 
            \multicolumn{2}{c}{Naive Reduction} &
            \multirow{2}{*}{Speedup} \\ \cmidrule{2-5}
        & {t ($\mu$s)} & {cyc/bfly} & {t ($\mu$s)} & {cyc/bfly}  \\ \midrule
        NTT (q)      & 57  & 7.34 & 393  & 50.59 & 6.9\\
        Inv. NTT (q) & 54  & 6.95 & 388  & 49.95 & 7.2\\ \midrule
        NTT (p)      & 43  & 5.54 & 240  & 30.89 & 5.6\\
        Inv. NTT (p) & 38  & 4.89 & 194  & 24.97 & 5.1\\ \bottomrule
    \end{tabular}
\end{table}

\begin{table}
\centering
    \caption{FHE Microbenchmarks}
    \begin{tabular}{cccccc} \toprule
        \multirow{2}{*}{Operation} & 
            \multicolumn{2}{c}{Fast Reduction} & 
            \multicolumn{2}{c}{Naive Reduction} &
            \multirow{2}{*}{Speedup} \\ \cmidrule{2-5}
        & {t ($\mu$s)} & {cyc/slot} & {t ($\mu$s)} & {cyc/slot}  \\ \midrule
        $\KeyGen$     & 232  & 328.5 &  952  & 1348.1 & 4.1  \\
        $\Encrypt$    & 186  & 263.4 &  621  &  879.4 & 3.3  \\
        $\Decrypt$    & 125  & 177.0 &  513  &  726.4 & 4.1  \\ \midrule
        $\SIMDadd$    &   5  &   8.1 &  393  &   49.7 & 6.1  \\
        $\SIMDmult$   &  10  &  14.7 &  388  &  167.1 & 11.3 \\ \midrule
        $\PermKeyGen$ & 466  & 659.9 & 1814  & 2568.7 & 3.9  \\
        $\Perm$       & 268  & 379.5 & 1740  & 2463.9 & 6.5  \\ 
        $\PermDecomp$ & 231  & 327.1 & 1595  & 2258.5 & 6.9  \\ 
        $\PermAuto$   &  35  &  49.6 &  141  &  199.7 & 4.0  \\ \bottomrule
    \end{tabular}
    \label{tab:micro-he}
\end{table}

Table \ref{tab:micro-perm} demonstrates the noise performance trade-off inherent in the permutation operation. Note that an individual permutation after the initial decomposition is roughly 8-9$\times$ faster than a permutation without any pre-computation. Finally we observe a linear growth in the run-time of the permutation operation with an increase in the number of windows, allowing us to trade off noise performance for run-time if few future operations are desired on the permuted ciphertext.

\begin{table}
\centering
    \caption{Permutation Microbenchmarks}
    \begin{tabular}{c c c c c} \toprule
        \multirow{2}{*}{\# windows} & {$\PermKeyGen$} & Key Size & {$\PermAuto$} & {Noise} \\ \cmidrule{2-5}
        & {t ($\mu$s)} & kB & {t ($\mu$s)} & bits  \\ \midrule
        3  &  466 &  49.15 & 35  & 29.3  \\
        6  &  925 &  98.30 & 57  & 19.3  \\
        12 & 1849 & 196.61 & 100 & 14.8  \\ \bottomrule
    \end{tabular}
    \label{tab:micro-perm}
\end{table}

\subsubsection{Linear Algebra Benchmarks}

Next we present micro-benchmarks for the linear algebra kernels. In particular we focus on matrix-vector products and 2D convolutions since these are the operations most frequently used in neural network inference. Before performing these operations, the server must perform a one-time {\em client-independent setup} that pre-processes the matrix and filter coefficients. In contrast with the offline phase of 2PC, this computation is NOT repeated per classification or per client and can be performed without any knowledge of the client keys. In the following results, we represent the time spent in this amortizable setup operation as $\mathsf{t_{setup}}$. Note that $\mathsf{t_{offline}}$ for both these protocols is zero. 

The matrix-vector product that we are interested in corresponds to the multiplication of a plaintext matrix with a packed ciphertext vector. We first start with a comparison of three matrix-vector multiplication techniques:
\begin{enumerate}
    \item \textbf{Naive}: Every slot of the output is generated independently by computing an inner-product of a row of the matrix with ciphertext column vector.
    \item \textbf{Diagonal}: Rotations of the input are multiplied by the generalized diagonals from the plaintext matrix and added to generate a packed output.
    \item \textbf{Hybrid}: Use the diagonal approach to generate a single output ciphertext with copies of the output partial sums. Use the naive approach to generate the final output from this single ciphertext
\end{enumerate}

We compare these techniques for the following matrix sizes: $2048\times 1$, $1024\times 128$, $128\times 16$. For all these methods we report the online computation time and the time required to setup the scheme in milliseconds. Note that this setup needs to be done exactly once per network and need not be repeated per inference. The naive scheme uses a 20bit plaintext window ($\mathsf{w_{pt}}$) while the diagonal and hybrid schemes use 10bit plaintext windows. All schemes use a 7bit relinearization window ($\mathsf{w_{relin}}$).

\begin{table}
\centering
    \caption{Matrix Multiplication Microbenchmarks}
\begin{tabular}{c c c c c c c}
\toprule
& & $\mathsf{\#in\_rot}$ & $\mathsf{\#out\_rot}$ 
    & $\mathsf{\#mac}$
    & $\mathsf{t_{online}}$ &  $\mathsf{t_{setup}}$ \\ \midrule
 \multirow{3}{*}{2048$\times$1}  & \textbf{N} & 0 & 11 & 1 & 7.9 & 16.1 \\ \cmidrule{2-7}
                                 & \textbf{D} & 2047 & 0 & 2048 & 383.3 & 3326.8 \\ \cmidrule{2-7}
                                 & \textbf{H} & 0 & 11 & 1 & 8.0 & 16.2 \\ \midrule
 \multirow{3}{*}{1024$\times$128}& \textbf{N} & 0 & 1280 & 128 & 880.0 & 1849.2 \\ \cmidrule{2-7}
                                 & \textbf{D} & 1023 & 1024 & 2048 & 192.4 & 1662.8 \\ \cmidrule{2-7}
                                 & \textbf{H} & 63 & 4 & 64 & 16.2 & 108.5 \\ \midrule
 \multirow{3}{*}{1024$\times$16} & \textbf{N} & 0 & 160 & 16 & 110.3 & 231.4 \\ \cmidrule{2-7}
                                 & \textbf{D} & 1023 & 1024 & 2048 & 192.4 & 1662.8 \\ \cmidrule{2-7}
                                 & \textbf{H} & 7 & 7 & 8 & 7.8 & 21.8 \\ \midrule
 \multirow{3}{*}{128$\times$16}  & \textbf{N} & 0 & 112 & 16 & 77.4 & 162.5 \\ \cmidrule{2-7}
                                 & \textbf{D} & 127 & 128 & 2048 & 25.4 & 206.8 \\ \cmidrule{2-7}
                                 & \textbf{H} & 0 & 7 & 1 & 5.3 & 10.5 \\ \bottomrule
\end{tabular}
\label{tab:micro-mm}
\end{table}

As seen in Section~\ref{sec:matmult} the online time for the matrix multiplication operation can be improved further by a judicious selection of the window sizes based on the size of the matrix used. Table \ref{tab:micro-mm-opt} shows the potential speed up possible from optimal window sizing. Note that although this optimal choice reduces the online run-time, the relinearization keys for all the window sizes must be sent to the server in the initial setup phase.

\begin{table}
\centering
\caption{Hybrid Matrix Multiplication Window Sizing}
\label{tab:micro-mm-opt}
\begin{tabular}{c c c c c c c}
\toprule
& $\mathsf{w_{pt}}$ & $\mathsf{w_{relin}}$ & $\mathsf{t_{online}}$
    & Speedup &  $\mathsf{t_{setup}}$ & Speedup \\ \midrule
2048$\times$1   & 20 & 20 &  3.6 & 2.2 &  5.7 & 2.9 \\
1024$\times$128 & 10 &  9 & 14.2 & 1.1 & 87.2 & 1.2 \\
1024$\times$16  & 10 &  7 &  7.8 & 1.0 & 21.5 & 1.0 \\
128$\times$16   & 20 & 20 &  2.5 & 2.1 &  3.7 & 2.8 \\ \bottomrule
\end{tabular}
\end{table}

Finally we remark that our matrix multiplication scheme is extremely parsimonious in the online bandwidth. The two-way online message sizes for all the matrices are given by $(w+1)*\mathsf{ct_{sz}}$ where $\mathsf{ct_{sz}}$ is the size of a single ciphertext (32 kB for our parameters).

Next we compare the two techniques we presented for 2D convolution: input rotation \textbf{(I)} and output rotation \textbf{(O)} in Table~\ref{tab:micro-conv2d}. We present results for four convolution sizes with increasing complexity. Note that the $5\times 5$ convolution is strided convolution with a stride of 2. All results are presented with a 10bit $\mathsf{w_{pt}}$ and a 8bit $\mathsf{w_{relin}}$.

\begin{table}
\centering
    \caption{Convolution Microbenchmarks}
\begin{tabular}{c c c c c}
\toprule
Input & Filter & Algorithm & $\mathsf{t_{online}}$ &  $\mathsf{t_{setup}}$ \\ 
(W$\times$H, C) & (W$\times$H, C) &  & \text{(ms)} &  \text{(ms)} \\ \midrule
\multirow{2}{*}{$(28\times28, 1)$}   & \multirow{2}{*}{$(5\times5, 5)$}   & \textbf{\text{I}} & 14.4 & 11.7 \\
                                  &                                 & \textbf{\text{O}} & 9.2  & 11.4 \\ \midrule
\multirow{2}{*}{$(16\times16, 128)$} & \multirow{2}{*}{$(1\times1,128)$} & \textbf{\text{I}} & 107  & 334 \\
                                  &                                 & \textbf{\text{O}} & 110  & 226 \\ \midrule
\multirow{2}{*}{$(32\times32, 32)$}  & \multirow{2}{*}{$(3\times3,32)$}  & \textbf{\text{I}} & 208  & 704 \\
                                  &                                 & \textbf{\text{O}} & 195  & 704 \\ \midrule
\multirow{2}{*}{$(16\times16, 128)$} & \multirow{2}{*}{$(3\times3,128)$} & \textbf{\text{I}} & 767  & 3202 \\
                                  &                                 & \textbf{\text{O}} & 704  & 3312 \\ \bottomrule
\end{tabular}
\label{tab:micro-conv2d}
\end{table}

As seen from Table~\ref{tab:micro-conv2d}, the output rotation variant is usually the faster variant since it reuses the same input multiple times. Larger filter sizes allow us to save more rotations and hence experience a higher speed-up, while for the $1\times1$ case the input rotation variant is faster. Finally, we note that in all cases we pack both the input and output activations using the minimal number of ciphertexts.

\subsubsection{Square, ReLU and MaxPool Benchmarks}

We round our discussion of the operation micro-benchmarks with the various activation functions we consider. In the networks of interest, we come across two major activation functions: Square and ReLU. Additionally we also benchmark the MaxPool layer with $(2\times2)$-sized windows. 

For square pooling, we implement a simple interactive protocol using our additively homomorphic encryption scheme. For ReLU and MaxPool, we implement a garbled circuit based interactive protocol. The results for both are presented in Table~\ref{tab:micro-activation}.

\begin{table}
\centering
    \caption{Activation and Pooling Microbenchmarks}
\begin{tabular}{c c c c c c}
\toprule
\multirow{2}{*}{Algorithm} & \multirow{2}{*}{Outputs} & $\mathsf{t_{offline}}$  &  $\mathsf{t_{online}}$ & $\mathsf{BW_{offline}}$ &  $\mathsf{BW_{online}}$\\
 &  & (ms)  & (ms) & (MB) &  (MB)\\ \midrule
Square & 2048 & 0.5 & 1.4 & 0 & 0.093 \\ \midrule
\multirow{2}{*}{ReLU} & 1000  & 89  & 201  & 5.43 & 1.68 \\
                      & 10000 & 551 & 1307 & 54.3 & 16.8 \\ \midrule
\multirow{2}{*}{MaxPool} & 1000  & 164 & 426 & 15.6 & 8.39 \\
                         & 10000 & 1413 & 3669 & 156.0 & 83.9 \\ \bottomrule
\end{tabular}
\label{tab:micro-activation}
\end{table}

\section{Network Benchmarks and Comparison}
\label{sec:networks}

Next we compose the individual layers from the previous sections and evaluate complete networks. For ease of comparison with previous approaches, we report runtimes and network bandwidth for MNIST and CIFAR-10 image classification tasks. We segment our comparison based on the CNN topology. This allows us to clearly demonstrate the speedup achieved by \gazelle\ as opposed to gains through network redesign.

\subsection{The MNIST Dataset.}

MNIST is a basic image classification task where we are provided with a set of $28\times28$ grayscale images of handwritten digits in the range $[0-9]$. Given an input image our goal is to predict the correct handwritten digit it represents. We evaluate this task using four published network topologies which use a combination of $\fc$ and $\conv$ layers:
\begin{enumerate}
    \item \textbf{A}: 3-$\fc$ layers with square activation from \cite{secureml}.
    \item \textbf{B}: 1-$\conv$ and 2-$\fc$ layers with square activation from \cite{cryptonets}.
    \item \textbf{C}: 1-$\conv$ and 2-$\fc$ layers with ReLU activation from \cite{deepsecure}.
    \item \textbf{D}: 2-$\conv$ and 2-$\fc$ layers with ReLU and MaxPool from \cite{minionn}.
\end{enumerate}

Runtime and the communication required for classifying a single image for these four networks are presented in table \ref{tab:mnist}.

\begin{table}
\centering
    \caption{MNIST Benchmark}
\resizebox{\linewidth}{!}{\begin{tabular}{c c c c c c c c}
\toprule
& \multirow{2}{*}{Framework} & \multicolumn{3}{c}{Runtime (s)} & \multicolumn{3}{c}{Communication (MB)} \\ \cmidrule{3-8}
 &  & Offline  & Online & Total & Offline & Online & Total \\ \midrule
\multirow{3}{*}{A} & SecureML   & 4.7 & 0.18 & 4.88 & - & -   & - \\
                   & MiniONN    & 0.9 & 0.14 & 1.04 & 3.8 & 12 & 47.6 \\
                   & \gazelle   & 0 & 0.03 & 0.03 & 0 & 0.5 & 0.5 \\ \midrule
\multirow{3}{*}{B} & CryptoNets & - & - & 297.5 & - & -   & 372.2 \\
                   & MiniONN    & 0.88 & 0.4 & 1.28 & 3.6 & 44 & 15.8 \\
                   & \gazelle   & 0 & 0.03 & 0.03 & 0 & 0.5 & 0.5 \\ \midrule
\multirow{3}{*}{C} & DeepSecure & - & - & 9.67 & - & -   & 791 \\
                   & Chameleon  & 1.34 & 1.36 & 2.7 & 7.8 & 5.1 & 12.9 \\
                   & \gazelle   & 0.15 & 0.05 & 0.20 & 5.9 & 2.1 & 8.0 \\ \midrule
\multirow{3}{*}{D} & MiniONN    & 3.58 & 5.74 & 9.32 & 20.9 & 636.6 & 657.5 \\
                   & ExPC       & - & - & 5.1 & - & - & 501 \\
                   & \gazelle   & 0.481 & 0.33 & 0.81 & 47.5 & 22.5 & 70.0 \\ \bottomrule
\end{tabular}}
\label{tab:mnist}
\end{table}

For all four networks we use a 10bit $\mathsf{w_{pt}}$ and a 9bit $\mathsf{w_{relin}}$.

Networks A and B use only the square activation function allowing us to use a much simpler AHE base interactive protocol, thus avoiding any use of GC's. As such we only need to transmit short ciphertexts in the online phase. Similarly our use of the AHE based $\fc$ and $\conv$ layers as opposed to multiplications triples results in 5-6$\times$ lower latency compared to \cite{minionn} and \cite{secureml} for network A. The comparison with \cite{cryptonets} is even more the stark. The use of AHE with interaction acting as an implicit bootstraping stage allows for aggressive parameter selection for the lattice based scheme. This results in over 3 orders of magnitude savings in both the latency and the network bandwidth.

Networks C and D use ReLU and MaxPool functions which we implement using GC. However even for these the network our efficient $\fc$ and $\conv$ implementation allows us roughly 30$\times$ and 17$\times$ lower runtime when compared with \cite{chameleon} and \cite{minionn} respectively. Furthermore we note that unlike \cite{chameleon} our solution does not rely on a trusted third party.

\subsection{The CIFAR-10 Dataset.}

The CIFAR-10 task is a second commonly used image classification benchmark that is substantially more complicated than the MNIST classification task. The task consists of classifying $32\times32$ color with 3 color channels into 10 classes such as automobiles, birds, cats, etc. For this task we replicate the network topology from \cite{minionn} to offer a fair comparison. We use a 10bit $\mathsf{w_{pt}}$ and a 8bit $\mathsf{w_{relin}}$.

\begin{table}
\centering
    \caption{CIFAR-10 Benchmark}
\resizebox{\linewidth}{!}{\begin{tabular}{c c c c c c c c}
\toprule
& \multirow{2}{*}{Framework} & \multicolumn{3}{c}{Runtime (s)} & \multicolumn{3}{c}{Communication (MB)} \\ \cmidrule{3-8}
 &  & Offline  & Online & Total & Offline & Online & Total \\ \midrule
\multirow{2}{*}{A} & MiniONN    & 472 & 72 & 544 & 3046 & 6226 & 9272 \\
                   & \gazelle   & 9.34 & 3.56 & 12.9 & 940 & 296 & 1236 \\ \midrule
\end{tabular}}
\label{tab:cifar10}
\end{table}

We note that the complexity of this network when measure by the number of multiplications is $~500\times$ that used in the MNIST network from \cite{deepsecure}, \cite{chameleon}. By avoiding the need for multiplication triples \gazelle\ offers a $50\times$ faster offline phase and a $20\times$ lower latency per inference showing that our results from the smaller MNIST networks scale to larger networks.

\section{Conclusions and Future Work}
\label{sec:conclusions}

In conclusion, this work presents \gazelle, a low-latency framework for secure neural network inference. \gazelle\ uses a judicious combination of packed additively homomorphic encryption and garbled circuit based two-party computation to obtain $20-30\times$ lower latency and $2.5-88\times$ lower online bandwidth when compared with multiple two-party computation based state-of-art secure network inference solutions~\cite{minionn,secureml,chameleon,deepsecure}, and more than 3 orders of magnitude lower latency and 2 orders of magnitude lower bandwidth than purely homomorphic approaches~\cite{cryptonets}. We briefly recap the key contributions of our work that enable this improved performance:
\begin{enumerate}
    \item Selection of prime moduli that simultaneously allow single instruction multiple data (SIMD) operations, low noise growth and division-free and lazy modular reduction.
    \item Avoidance of ciphertext-ciphertext multiplications to reduce noise growth.
    \item Use of secret-sharing and interaction to emulate a lightweight bootstrapping procedure allowing for the composition of multiple layers to evaluate deep networks.
    \item Homomorphic linear algebra kernels that make efficient use of the automorphism structure enabled by a power-of-two slot-size.
    \item Sparing use of garbled circuits limited to ReLU and MaxPooling non-linearities that require linear-sized Boolean circuits.
    \item A compact garbled circuit-based transformation gadget that allows to securely compose the  $\pahename$-based and garbled circuit based layers.
\end{enumerate}

We envision the following avenues to extend our work on \gazelle\ and make it more broadly applicable. A natural next step is to handle larger application-specific neural networks that work with substantially larger inputs to tackle data analytics problems in the medical and financial domains. In ongoing work, we extend our techniques to a large variety of classic two-party tasks such as privacy-preserving face recognition \cite{SSW09} which can be factored into linear and non-linear phases of computation similar to what is done in this work. In the low-latency LAN setting, it would also be interesting to evaluate the impact of switching out the garbled-circuit based approach for a GMW-based approach which would allow us to trade off latency to substantially reduce the online and offline bandwdith. A final, very interesting and ambitious line of work would be to build a compiler that allows us to easily express arbitrary computations and automatically factor the computation into $\pahename$ and two-party primitives. 

\ifnum\anon=0
\subsubsection*{\sc acknowledgments} We thank Kurt Rohloff, Yuriy Polyakov and the PALISADE team for providing us with access to the PALISADE library. We thank Shafi Goldwasser, Rina Shainski and Alon Kaufman for delightful discussions. We thank our sponsors, the Qualcomm Innovation Fellowship and Delta Electronics for supporting this work.
\fi

\bibliographystyle{plain}
\bibliography{References}

\end{document}